\documentclass[reprint,amsmath,amssymb,aps,pre,%
  superscriptaddress,a4paper,footinbib]{revtex4-2}
\usepackage{newtxtext,newtxmath}
\usepackage{microtype}
\usepackage{gensymb}
\usepackage[colorlinks=true,allcolors = blue]{hyperref}
\usepackage{xcolor}
\usepackage{graphicx}
\usepackage{dcolumn}
\usepackage{bm}
\usepackage{cases}
\usepackage{siunitx}

\usepackage[version=3]{mhchem} 

\newcommand{\um}{\upmu\mathrm{m}}
\newcommand{\persec}{\mathrm{s}^{-1}}

\newcommand{\textsumi}{{\textstyle\sum_i}}

\newcommand{\kB}{k_{\mathrm{B}}}
\newcommand{\kT}{\kB T}

\newcommand{\myvec}[1]{{\mathbf{#1}}}
\newcommand{\Ivec}{\myvec{I}}
\newcommand{\Evec}{\myvec{E}}
\newcommand{\Uvec}{\myvec{U}}
\newcommand{\Avec}{\myvec{A}}
\newcommand{\Bvec}{\myvec{B}}
\newcommand{\Hvec}{\myvec{H}}
\newcommand{\Jvec}{\myvec{J}}
\newcommand{\Rvec}{\myvec{R}}

\newcommand{\jvec}{\myvec{j}}

\newcommand{\Svec}{\myvec{S}}
\newcommand{\nvec}{\myvec{n}}
\newcommand{\rvec}{\myvec{r}}
\newcommand{\tvec}{\myvec{t}}

\newcommand{\vvec}{\myvec{v}}
\newcommand{\bvec}{\myvec{b}}

\renewcommand{\div}{\nabla\cdot} 
\newcommand{\grad}{\nabla}
\newcommand{\curl}{\nabla\times}
\newcommand{\delsq}{\nabla^2}

\newcommand{\ehat}{\hat{\myvec{e}}}
\newcommand{\rhat}{\hat{\rvec}}

\newcommand{\lD}{\lambda_{\text{D}}}

\newcommand{\dl}{\mathrm{d}\myvec{l}}

\newcommand{\dA}{\mathrm{d}{A}}
\newcommand{\dAvec}{\mathrm{d}\myvec{A}}
\newcommand{\ds}{\mathrm{d}{s}}
\newcommand{\dV}{\mathrm{d}{V}}

\newcommand{\vslip}{\vvec^{\text{s}}}

\newcommand{\Eq}[1]{Eq.~\eqref{#1}}
\newcommand{\Eqs}[1]{Eqs.~\eqref{#1}}
\newcommand{\Fig}[1]{Fig.~\ref{#1}}
\newcommand{\Figs}[1]{Figs.~\ref{#1}}
\newcommand{\Table}[1]{Table~\ref{#1}}
\newcommand{\Refcite}[1]{Ref.~[\onlinecite{#1}]}
\newcommand{\Refscite}[1]{Refs.~[\onlinecite{#1}]}
\newcommand{\Appendix}[1]{Appendix~\ref{#1}}

\newcommand{\partf}[3]{{#1}~\hyperref[#2]{\ref*{#2}#3}}
\newcommand{\partFig}[2]{\partf{Fig.}{#1}{#2}}

\newcommand{\latin}[1]{{\itshape #1}}
\newcommand{\eg}{\latin{e.\,g.}}
\newcommand{\ie}{\latin{i.\,e.}}
\newcommand{\etal}{\latin{et al.}}
\newcommand{\etc}{\latin{etc}}
\newcommand{\perse}{\latin{per se}}
\newcommand{\via}{\latin{via}}
\newcommand{\confer}{\latin{cf.}}
\newcommand{\modulo}{\latin{modulo}}

\begin{document}

\title{Salt solutions with two or more salts generate ion currents analogous to magnetic field lines}

\author{Patrick B. Warren}
\email{patrick.warren@stfc.ac.uk}
\affiliation{STFC Hartree Centre, Sci-Tech Daresbury, Warrington, WA4 4AD, United Kingdom}

\author{Richard P. Sear}
\email{r.sear@surrey.ac.uk}
\homepage{https://richardsear.me/}
\affiliation{School of Mathematics and Physics, University of Surrey, Guildford, GU2 7XH, United Kingdom}

\date{\today}

\begin{abstract}
A gradient of a single salt in a solution generates an electric field, but not a current. Recent theoretical work by one of us [Phys.\ Rev.\ Lett.\ {\bf24}, 248004 (2020)] showed that the Nernst-Planck equations imply that crossed gradients of two or more different salts generate ion currents. These currents in solution have associated non-local electric fields.  Particle motion driven by these non-local fields has recently been observed in experiment by Williams \etal~[Phys.\ Rev.\ Fluids {\bf9}, 014201 (2024)]\,; a phenomenon which was dubbed action-at-a-distance diffusiophoresis. Here we use a magnetostatic analogy to show that in the far-field limit, these non-local currents and electric fields both have the functional form of the magnetic field of a magnetic dipole, decaying as $r^{-d}$ in $d=2$ and $d=3$ dimensions. These long-ranged electric fields are generated entirely within solutions and have potential practical applications since they can drive both electrophoretic motion of particles, and electro-osmotic flows. The magnetostatic analogy also allows us to import tools and ideas from classical electromagnetism, into the study of multicomponent salt solutions.
\end{abstract}

\maketitle

\section{Introduction}
It has been known since the end of the 19th century that whenever salts diffuse, electric fields are generated in the solution \cite{probstein1994, rieger1994, newman2004}.  This is described by the Nernst-Planck (NP) equations \cite{probstein1994}.  For a single binary electrolyte (\ie\ a single anion and single cation), the electric field obtained by injecting charge neutrality into the NP equations is determined only by the \emph{local} salt concentration gradient.  However, if there are multiple salts present, the situation is not as straightforward. Unless the gradients of the different salts are all parallel, the charge-neutral NP equations predict a \emph{non-local} contribution to the electric field \cite{warren2020, williams2024}, \ie\ the electric field at a point is not determined solely by the ion concentration gradients at that point.  These non-local electric fields correspond to autonomous ion currents in solution. Here, autonomous means that they arise internally from the salt concentration gradients, and are not generated electrochemically by injecting or removing ions using electrodes, as is done in electrochemical cells.  Both the field and current spread out through the solution into regions beyond the source concentration gradients. The electric fields cause electrophoresis of suspended particles in such regions, which was dubbed `action-at-a-distance' diffusiophoresis~\cite{williams2024}.  We expect this phenomenology of non-local fields and ion currents to be ubiquitous in systems where there are multiple salts and non-aligned concentration gradients.  

\begin{figure}[b]
\includegraphics[width=0.8\columnwidth]{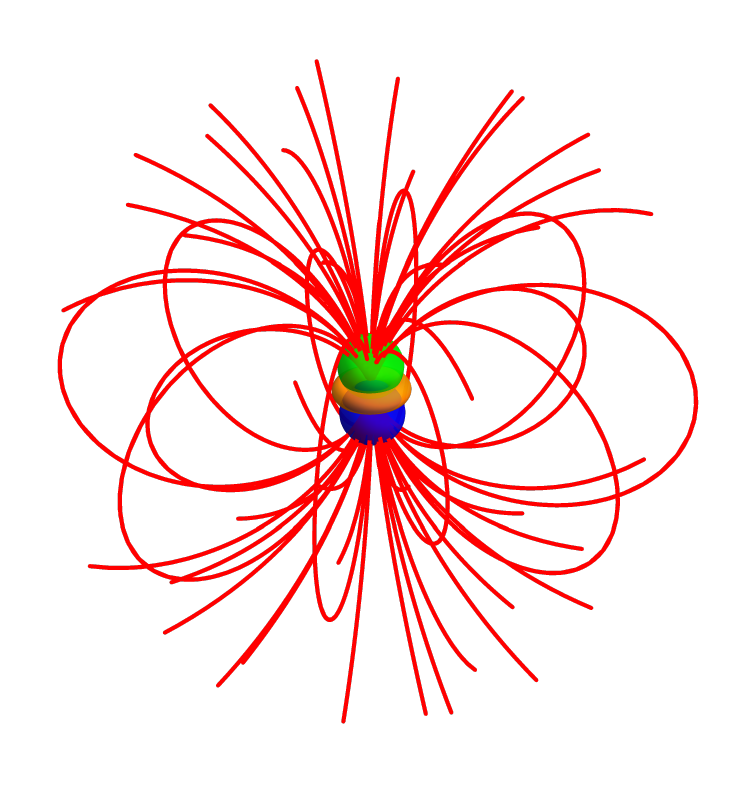}
\caption{Magnetostatic analogy in the Nernst-Planck equations: two salt sources (blue and green) overlap in the doughnut-shape region (orange), generating circulating ion currents (red lines).  These are analagous to the magnetic field lines around a small current loop.\label{fig:farfield3d}}
\end{figure}

Williams \etal~\cite{williams2024} recently observed non-local diffusiophoresis in experiments, in a quasi-two-dimensional system where the salt gradients were generated by so-called soluto-inertial beacons~\cite{banerjee2019}.  With a suitable choice of salts to minimise direct diffusiophoresis for one component, the non-local effect could clearly be seen.  That work identified an unexpected and fascinating magnetostatic analogy for the charge-neutral NP equations. This analogy allows a greatly improved intuitive understanding of how and under what conditions autonomous ion currents, electric fields, and the associated non-local diffusiophoresis phenomenon arise.  
The focus of the present work is to explore this analogy. We use it to calculate ion currents in simple geometries, and to show, as illustrated in \Fig{fig:farfield3d}, that far from crossed gradients the currents have the same form as lines of force of a magnetic dipole.

\begin{table}
\begin{ruledtabular}
  \begin{tabular}{lllll}
    \multicolumn{2}{c}{Nernst-Planck} &
    \multicolumn{3}{c}{Magnetostatics} \\
    \hline\\[-6pt]
    & $\div\Ivec = 0$ && $\div\Bvec = 0$  & Gauss' law \\[3pt]
    & $\curl(\varrho\Ivec) = \Svec$ &&
    $\curl\Hvec=\Jvec$ & Amp\`ere's law \\[3pt]
    & $\div\Svec=0$ && $\div\Jvec=0$ & current conservation\\[3pt]
    & $\nvec\cdot\Ivec = 0$ && $\nvec\cdot\Bvec=0$ & boundary condition
\end{tabular}
\end{ruledtabular}
\caption{Magnetostatic analogy for the ion current in the Nernst-Planck equations, where $\Svec=\grad g\times\grad\varrho$. Note that $\Bvec=\mu_0\Hvec$.  The boundary condition on $\Bvec$ is a perfect conductor.  The magnetostatics notation is from Jackson's {\it Classical Electrodynamics}, 3rd ed.\ \cite{jackson1999}.\label{tab:mag}}
\end{table}

The paper is laid out as follows.  In section \ref{sec:NP}, starting from the NP equations, we derive expressions for the ion currents.  We first use this to derive a Poisson-like equation for the electrostatic potential. This leads to a useful uniqueness theorem~\cite{newman2004, warren2020}.  We next develop what we term the ion current equations and demonstrate the analogy to Gauss' and Amp\`ere's law in magnetostatics \cite{jackson1999, griffiths2013}. The analogous expressions are in \Table{tab:mag} (we explain the Nernst-Planck column notation below).  We also define a quantity analogous to the magnetic vector potential, and the associated vector potential equation. In section \ref{sec:2d} we focus on two-dimensional systems, and we show that the $z$-component of the vector potential obeys a pseudo-scalar Poisson equation.  This allows us to make statements about the far-field behaviour of the ion current in two-dimensional problems.  In a final section we explore the implications for diffusiophoresis of suspended particles \cite{Anderson1982, *Prieve1984, Anderson1989, shin2016, Velegol2016, Marbach2019}, and we close with general remarks about the practical developments and potential applications of the effects we describe.

\section{Nernst-Planck equations}\label{sec:NP}
The NP equations are a set of equations, one for each species of ion, which govern ion transport in an electrolyte.  The NP equation for the $i$-th ion is
\begin{equation}
  \frac{\partial c_i}{\partial t}+\div\jvec_i=0\,,\quad
  \jvec_i=-D_i(\grad c_i+z_i c_i\grad\varphi)\,.\label{eq:np}
\end{equation}
In this, $c_i$ is the concentration of the $i$-th ion, $\jvec_i$ is the associated flux, $D_i$ is a diffusion coefficient, $z_i$ is the valence in units of the elementary charge $e$, and $\varphi=e\phi/\kT$ is a non-dimensionalised form of the electrostatic potential $\phi$, where $\kB$ is Boltzmann's constant and $T$ is the temperature.  These should be complemented  (see \Appendix{app:neut}) by the charge neutrality constraint, $\sum_i z_i c_i = 0$.

The ion current in the NP equations is
\begin{equation}
  \Ivec=\textsumi\,z_i\,\jvec_i\,.\label{eq:isum}
\end{equation}
In cases where $\Ivec=0$, the potential gradient $\grad\varphi$ can be eliminated from the NP equations, which then reduce to a coupled diffusion problem \cite{newman2004, gupta2019}.  In general though the ion current does not vanish. However, if there are no sources or sinks of ions (no electrodes), it must be solenoidal
\begin{equation}
  \div\Ivec=0\,.\label{eq:divi}
\end{equation}
This follows directly if one assumes, as we do, that charge neutrality holds at all times.

Introducing the weighted sums
\begin{equation}
  g=\textsumi\,z_i D_i c_i\,,\quad
  \sigma=\textsumi\,z_i^2D_i c_i\,,\label{eq:sigmag}
\end{equation}
where $\sigma$ is the ionic conductivity, we find
\begin{equation}
  \Ivec=-\grad g-\sigma\grad\varphi\,.\label{eq:i1}
\end{equation}
This follows from injecting NP equations into \Eq{eq:isum}.

%
The NP equations govern the temporal and spatial evolution of the concentration fields $c_i(\rvec, t)$.  This happens on a time scale which is set by salt diffusion coefficients and the overall size of the system, typically seconds to minutes.  On the other hand as we argue in \Appendix{app:neut}, once salt gradients are given, the ion current and associated electric field relax on the Debye time  which is typically much less than a microsecond \cite{hafeman1965, bazant2004, rubinstein2009, barnaveli2024}.  Therefore, for all practical purposes, we can view the ion concentration fields as `frozen' on this time scale.  This justifies the implied steady state in \Eq{eq:i1}.  

\subsection{Potential equation}
Equation~\eqref{eq:i1} or its equivalent can be found in many textbooks~\cite{probstein1994, newman2004}.  If we combine it with $\div\Ivec=0$, we see that $\varphi$ obeys an inhomogeneous Poisson equation \cite{probstein1994, warren2020},
\begin{equation}
  \div(\sigma\grad\varphi)+\delsq g=0\,.\label{eq:pot}
\end{equation}
The boundary conditions are typically $\nvec\cdot\Ivec=0$ at a surface, where $\nvec$ is the surface normal, or $|\Ivec|\to0$ at large distances.  Here $\Ivec$ is expressed in terms of $\grad\varphi$ according to \Eq{eq:i1}, so that in terms of the inhomogeneous Poisson equation in \Eq{eq:pot} these boundary conditions are usually of the Neumann type.  With boundary conditions such as these, \Eq{eq:pot} determines $\varphi$ to within an additive constant.  The system is analagous to the problem of finding the electrostatic potential in a medium with an inhomogeneous permittivity and as such one can anticipate the solution will be unique.  An explicit proof for the present situation was given in \Refcite{warren2020}. 

\begin{figure}[t]
\includegraphics[width=0.9\columnwidth]{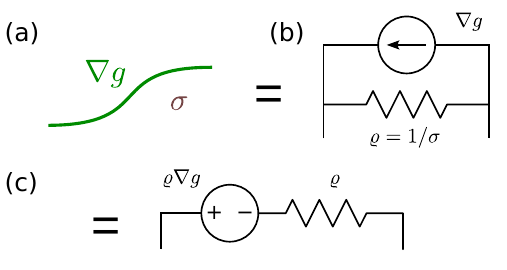}
\caption{Circuit element interpretation of \Eqs{eq:i1} and~\eqref{eq:i3}: (a) gradient $\nabla g$ in a background electrolyte can be interpreted as (b) an ideal current source in parallel with a resistance for \Eq{eq:i1}, or (c) an ideal voltage source in series with the same resistance for \Eq{eq:i3}.\label{fig:gradg}}
\end{figure}

\subsubsection{Equivalent circuit models}
We can use the language of  electrical circuit theory \cite{horowitz2015} to describe \Eq{eq:i1}.  The current in a region of the solution can be thought as due to two elements in parallel with each other (\Fig{fig:gradg}).  The first term in \Eq{eq:i1} is like an ideal current source, equal to $\grad g$.  The second term is then like an internal resistance verifying Ohm's law, in parallel with the current source.  This interpretation (\partFig{fig:gradg}{b}) of \Eq{eq:i1} sits alongside $\div\Ivec=0$, which is the equivalent in this context of Kirchoff's first law in electrical circuit theory; Kirchoff's second law corresponds to the fact that the electric field $\Evec=-\grad\varphi$ is conservative.

\begin{table}
\begin{ruledtabular}
  \begin{tabular}{ccc}
    {potential-conductivity picture} &&
    {current-resistivity picture} \\
    \hline\\[-6pt]
    $\div(\sigma\grad\varphi)+\delsq g=0$ &&
    $\curl(\varrho\Ivec) = \grad g\times\grad\varrho\,,\quad\div\Ivec=0$ \\[3pt]
    $\Ivec=-\grad g-\sigma\grad\varphi$ &&
    $-\grad\varphi=\varrho\Ivec+\varrho \grad g$ \\
\end{tabular}
\end{ruledtabular}
\caption{Dual but equivalent formulations for the electrostatic potential $\varphi$ and the ion current $\Ivec$ in the NP equations.  On the left-hand side the primary focus is on $\varphi$ in the inhomogeneous Poisson equation in \Eq{eq:pot}.  On the right-hand side the primary focus is on $\Ivec$ in the ion current equations.\label{tab:dual}}
\end{table}

Equations~\eqref{eq:i1} and~\eqref{eq:pot} form what we shall term the \emph{potential-conductivity} or $(\varphi,\sigma)$ picture (\Table{tab:dual}) for the origin of autonomous ion currents in the NP equations (with local charge neutrality).  From this point of view, the primary focus is on solving the inhomogeneous Poisson equation in \Eq{eq:pot} for the electrostatic potential $\varphi$, given $g$ and $\sigma$.  After this, the ion current $\Ivec$ follows from \Eq{eq:i1}.

\subsection{Ion current equations}\label{sec:ion_current}
We now turn to an alternate way of specifying the problem, which leads to the magnetostatic analogy described in the introduction.  We start by rearranging \Eq{eq:i1} to express the electric field $\Evec=-\grad\varphi$ in terms of the ion current $\Ivec$, the resistivity $\varrho=1/\sigma$, and $\grad g$,
\begin{equation}
  \Evec=\varrho\Ivec+\varrho \grad g\,.\label{eq:i3}
\end{equation}
The first term here is just Ohm's law again, and the second term is the contribution from the current source, here appearing in the guise of a diffuse liquid junction potential \cite{rieger1994, newman2004}.  \Eq{eq:i3} indicates that the previous circuit element model can also be interpreted as an ideal voltage source in series with the same internal resistance, shown in \partFig{fig:gradg}{c}.  This is the well-known Norton-Th\'evenin equivalence \cite{horowitz2015}.

Since $\curl\Evec=0$, one has \cite{williams2024}
\begin{equation}
  \curl(\varrho\Ivec)=\grad g\times \grad\varrho\,.\label{eq:current}
\end{equation}
The right-hand side derives from the vector calculus identity $\curl(\varrho\grad g) = -\grad g\times\grad\varrho$.  We call \Eq{eq:current} in combination with $\div\Ivec=0$ the \emph{ion current equations}.  They should be combined with the boundary conditions on $\Ivec$ to fully specify the problem.

Equations~\eqref{eq:i3} and~\eqref{eq:current} with $\div\Ivec=0$ form what we shall call the \emph{current-resistivity} or $(\Ivec,\varrho)$ picture (\Table{tab:dual}), for the origin of the ion currents in the NP equations.  This is a dual to the above potential-conductivity picture.  From this point of view, the primary focus is on solving \Eq{eq:current} for $\Ivec$, given $g$ and $\varrho$, supplemented by $\div\Ivec=0$.  After this has been obtained, the electric field follows from \Eq{eq:i3} and the potential can be found by integrating $\Evec=-\grad\varphi$.

\begin{figure}[t]
\includegraphics[width=0.9\columnwidth]{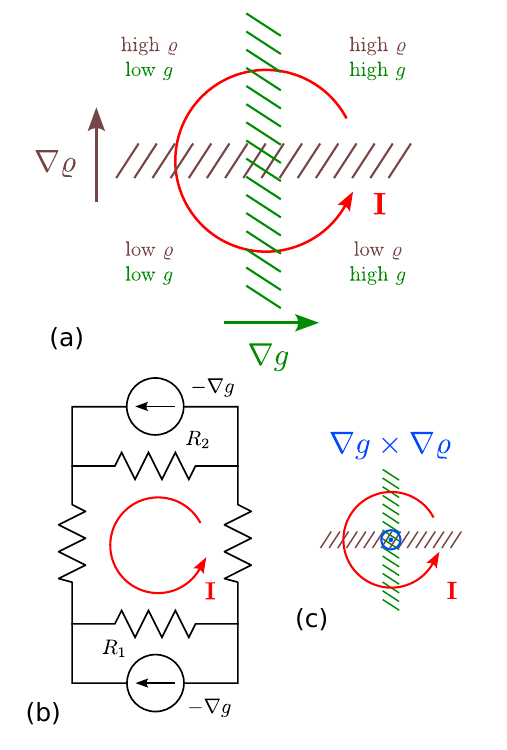}
\caption{Crossed gradients in $g$ and $\varrho$ generate an ion current which circulates anticlockwise as indicated in (a).  In an equivalent circuit model, shown in (b), the current circulates this way because $R_2>R_1$.  From the point of view of crossed gradients shown in (c), $\grad g\times \grad\varrho$ points out of the page and the direction of circulation is determined by the right-hand rule as in Amp\`ere's law in magnetostatics.\label{fig:crossgrad}}
\end{figure}

To complete the picture, we should address the question of whether $\Ivec$ is uniquely determined by the ion current equations.  Note that we cannot appeal to the Helmholtz theorem which states that (\modulo\ some reasonable conditions) a vector field is uniquely determined by its curl and divergence \cite{griffiths2013} because in the present case two different vector fields are involved, namely $\varrho\Ivec$ and $\Ivec$ since we need to allow for the fact that $\varrho$ is spatially varying.  We can answer the posed question in the affirmative though by utilising the fact that $\grad g\times\grad\varrho = \curl(-\varrho\grad g)$.  Then, \Eq{eq:current} can be written as $\curl(\varrho\Ivec+\varrho\grad g)=0$.  This can be integrated to get $\varrho\Ivec+\varrho\grad g =- \grad\varphi$, where $\varphi$ is an as-yet-undetermined scalar.  If we now impose $\div\Ivec=0$, we deduce that $\varphi$ solves \Eq{eq:pot}.  But as we have claimed, and is proven in \Refcite{warren2020}, this determines $\varphi$ uniquely up to a constant.  The constant vanishes in taking the gradient in forming $\Ivec=-\grad g -\sigma \grad\varphi$, so we conclude that $\Ivec$ is indeed uniquely determined in the current-resistivity picture.  

\subsection{Comparison between the two pictures}
The two points of view just described interpret the origins of the autonomous ion current in the NP equations from markedly different perspectives, although they are describing the exact same underlying physics.  The differences can be illustrated by considering the simplest case of crossed gradients in $g$ and $\varrho$, shown in \Fig{fig:crossgrad}. 

From the point of view of the potential-conductivity picture, the gradient in $g$ corresponds to a current source which is everywhere the same, but is more effective at injecting an ion current in the upper half-plane, since the internal resistance is higher there (lower conductivity).  This can be seen by considering the equivalent circuit model \cite{norton-note} shown in \partFig{fig:crossgrad}{b}.  Here, the current sources attempt to drive a current in opposite directions around the loop, but it is relatively easy to show that as long as $R_2>R_1$ (the other resistances are irrelevant), the net circulation is anticlockwise as indicated.

On the other hand, the current-resistivity picture gives a much simpler indication of the direction in which the ion current circulates (\partFig{fig:crossgrad}{c}).  According to \Eq{eq:current}, the crossed gradients represent a source which is localised in the centre, where the gradients in $g$ and $\varrho$ overlap.  The right-hand rule then indicates the direction of circulation of the ion current, which is of course in accord with the previous description.  This latter interpretation closely resembles the magnetic field lines which circulate around an infinite steady line current.  As we shall see in the next section, this is more than just a coincidence since the analogy with magnetic fields generated by steady electrical currents (magnetostatics) can be placed on a formal footing, although it is somewhat imperfect.

\subsection{Magnetostatic analogy}\label{subsec:magneto}
We are now in a position to set out the magnetostatic analogy in the problem, summarised in \Table{tab:mag}.  For the magnetostatic sector, we use the terminology and notation from Jackson~\cite{jackson1999}.

In the analogy, the ion current $\Ivec$ maps to the magnetic flux density $\Bvec$: both have zero divergence --- Gauss' law; and the product $\varrho\Ivec$ maps to the magnetic field intensity $\Hvec$ --- Amp\`ere's law. The source of $\Hvec$ in magnetostatics is the electrical current density $\Jvec$ (this should not to be confused with the ion fluxes in the NP equations); we can make the analogy even sharper (\Table{tab:mag}) by writing our analog of Amp\'{e}re's law as
\begin{subequations}
\begin{equation}
    \curl(\varrho\Ivec)= \Svec
\end{equation}
with
\begin{equation}
    \Svec =\grad g\times\grad\varrho\,.\label{eq:source}
\end{equation}
\end{subequations}
It follows from standard vector calculus identities \cite{jackson1999} that $\div\Svec=0$.  This corresponds to the conservation law $\div\Jvec=0$ for the electrical current in steady-state magnetostatics.

With this we can now see how an atonomous ion current develops in the NP equations in terms of the crossed gradient source $\Svec$.  An figurative example was already shown in \Fig{fig:farfield3d}.  Here, we envisage two dissolving salt sources or soluto-inertial beacons suspended in an electrolyte solution, generating diffuse radial gradients of ions shown in green and blue.  There should be three or more different species of ions, so one can think of for example dissolving NaCl and KCl crystals (\ie\ with a common anion), or the soluto-inertial beacons used by Williams \etal~\cite{williams2024}.  We shall assume a background electrolyte to provide a supporting conducting medium.  We suppose the sources are slightly displaced from each other, so that the gradients cross in a doughnut-shape region shown in orange where $|\Svec|>0$.  This region acts as a source for the ion current analogous to the canonical problem in magnetostatics of the magnetic field generated by a small current loop.  We use this analogy to draw the ion current lines in red.  These are simply the magnetic flux lines of a magnetic dipole centered on the $\Svec$-loop.  This will not be accurate near the sources, but we expect it to be correct at large distances, with the expectation that $|\Ivec|\sim r^{-3}$ for $r\to\infty$.  This shows that the ion current penetrates far into the surrounding region around the localised salt gradients, and consequently one would expect suspended particles in this region to show an electrophoretic drift in the associated electric field.  Since the whole phenomenon is driven by the salt gradients localised near the origin, this is a prototypical example of action-at-a-distance diffusiophoresis.

We should point out an imperfection exists in the above analogy, which is that the resistivity $\varrho$, which plays the role of the magnetic permeability (\Table{tab:mag}), appears on \emph{both} sides of \Eq{eq:current}.  In magnetostatics, it is natural to decouple the permeability from the electric current sources, although in reality the electric current has to be carried by a conductor, whose magnetic properties will differ in general from the bulk medium.  In the NP equations though, there can be no ion current without a gradient in resistivity to contribute to $\Svec\ne0$, and therefore the analogy has to be approached with a little care.  For example, as we have already seen in \Fig{fig:crossgrad} and as we shall explore further below, it is impossible to construct a situation which exactly corresponds to the canonical single isolated line current in magnetostatics, with circular magnetic flux lines; rather, the nearest one can do is either accept a resistivity gradient that persists to far field as in \partFig{fig:crossgrad}{a}, or consider a dipole line source, as a pair of equal and opposite $\Svec$ line sources in parallel (section \ref{subsec:ff2d}).

Since our ion current $\Ivec$ is solenoidal ($\div\Ivec=0$), we can write it as the curl of a vector potential, $\Ivec = \grad \times \Avec$.  Injecting this into \Eq{eq:current} obtains
\begin{equation}
  \curl[\varrho\,(\grad \times \Avec)] = \Svec\,.\label{eq:curl2}
\end{equation}
With this, we no longer need to separately consider the conservation law $\div\Ivec=0$.  This vector potential approach is most useful in considering two-dimensional systems, which we explore below in section \ref{sec:2d}.

The interested reader may wonder what is the analog of the Biot-Savart (BS) law in this problem.  The presence of $\varrho$ on both sides of \Eq{eq:current} lead to complications here, and the final result requires a generalisation of the familiar BS law in magnetostatics.  The issue is addressed in \Appendix{app:BS}.

\subsection{Far field of a localised source}\label{subsec:ff3d}
We now consider a localised source $\Svec$ in the ion current equations, such as the one in \Fig{fig:farfield3d}.  In this limit we can assume that the background resistivity $\varrho$ is constant, and therefore in far field, from \Eq{eq:current}, the ion current is irrotational ($\curl\Ivec=0$) as well as solenoidal ($\div\Ivec=0$).  In such a case we can write the ion current as the gradient of a scalar, which must be harmonic,
\begin{equation}
  \varrho\Ivec=-\grad\omega\,,\quad\delsq\omega=0\qquad\text{(far field)}\,.
\end{equation}
This maps the problem onto the far-field multipole expansion of the electrostatic potential around a localised charge distribution~\cite{jackson1999}.  Making use of this analogy we can write
\begin{equation}
  \omega=\frac{a}{4\pi r}+\frac{\bvec\cdot\rvec}{4\pi r^3}
  +\cdots\,,\label{eq:ffharm}
\end{equation}
where $r$ is the distance from the origin, and $a$, $\bvec$, \etc\ are coefficients.

We can show that the leading order (monopole) term in \Eq{eq:ffharm} vanishes, in other words $a=0$. Consider a large sphere $\cal S$ centered on the origin, and the integral of $\varrho\Ivec=-\grad\omega$ over the surface of this sphere, which is in the far field.  Then  $a=\varrho \int\! I_r\,\dA$, where $I_r$ is the radial component of $\Ivec$.  This follows by treating \Eq{eq:ffharm} as an expansion in spherical harmonics \cite{mathews1964, jackson1999}\,; orthogonality of the higher-order terms to $Y_{00}=1/\sqrt{4\pi}$ then guarantees that only the monopole term contributes to the integral. However, $\int\! I_r\,\dA = \int\Ivec\cdot\nvec\,\dA = 0$ by virtue of the divergence theorem applied to $\div\Ivec=0$, so $a=0$ always.

Given this, the leading order term for the ion current in far field, is the dipole term
\begin{equation}
  \Ivec=\frac{3(\bvec\cdot\rhat)\,\rhat-\bvec}{4\pi\varrho r^3}+\cdots\,.\label{eq:iff}
\end{equation}
This justifies the claims made above that in far field the ion current around a localised source resembles the magnetic field lines around a small current loop, with $|\Ivec|\sim r^{-3}$ at large distances.  

Unfortunately one cannot easily obtain an exact expression for $\bvec$ in terms of the source $\Svec$, since as already explained the resistivity $\varrho$ is intimately involved in both sides of \Eq{eq:current} which is the analog of Amp\`ere's law in this problem.  The generalised BS law in \Appendix{app:BS} suggests that $\bvec$ should be roughly of the order the first moment of $\Svec$ though. 
Then, if the spatial extent of $\Svec$ is order $w$, we expect $|\Svec|\sim w^{-2}$ since it is the product of two gradients, noting that the dimensions of $g$ and $\varrho$ cancel, and hence $|\bvec|\sim w\times w^{-2}\times w^3$ where the factors correspond to the moment length scale $\sim w$, the magnitude $|\Svec|\sim w^{-2}$, and the volume of the source $\sim w^3$.  This would imply $|\bvec|\sim w^2$ and therefore $|\Ivec|\sim w^2/\varrho r^3$ from \Eq{eq:iff}.  

Furthermore, in far field there are no gradients and so the electric field $\Evec=\varrho\Ivec$.  Since we define $\Evec=-\grad\varphi$ where $\varphi=e\phi/\kT$, electric fields in our system have units of inverse length.  Restoring the full units, we therefore expect the electric field in the far field of a localised source to be dipolar with
\begin{equation}
  |\Evec|\sim (\kT/e)\times w^2/r^3\qquad(r\gg w)\,.\label{eq:eff}
\end{equation}
This result can also be obtained by straightforward dimensional analysis.  If we require that the electric field is proportional to $\kT/e$ since that sets the scale for all autonomous electrical phenonomena, and inversely proportional to $r^3$ in far field ($r\gg w$) from \Eq{eq:iff}, the missing factor must have units of length squared.  Since the size of the localised source $w$ is the only independent length scale, one recovers \Eq{eq:eff}.

\subsection{Further results and identities}
We reprise some results from our earlier works \cite{warren2020, williams2024}.  These will be useful in the sequel.

\subsubsection{The surface integral over the source is zero, for open surfaces with perimeters in the far field}\label{subsec:surf0}
We start by applying standard vector calculus identities to the source $\Svec$ of \Eq{eq:source} 
\begin{equation}
 \Svec= \grad g\times \grad\varrho
  =\curl(g\grad\varrho)
  =-\curl(\varrho\grad g)\,.
\end{equation}
Applying Stokes' theorem
\begin{equation}
  {\textstyle\int} \Svec \cdot\dAvec
  ={\textstyle\oint} g\grad\varrho\cdot\dl
  =-{\textstyle\oint}\varrho\grad g\cdot\dl\,.\label{eq:oint}
\end{equation}
In the far field where $\varrho$ and $g$ are constant the line integral is zero so
\begin{equation}
    {\textstyle\int}\Svec\cdot\dAvec=0\quad
    \text{(perimeter in far field)}\,.\label{eq:Sarea0}
\end{equation}

\subsubsection{Line integrals over loops in the far field are zero}\label{subsec:loop0}
As $\Evec=-\grad\varphi$, it follows from \Eq{eq:i3} that
\begin{equation}
  {\textstyle\oint}\varrho(\Ivec+\grad g)\cdot\dl=0\,,\label{eq:loop}
\end{equation}
around a closed path. In the far field $\varrho$ and $g$ are both constant. So for any loop entirely in the far field the integral over the second term is zero. Therefore
\begin{equation}
  {\textstyle\oint}\Ivec\cdot\dl=0\quad
    \text{(perimeter in far field)}\,.\label{eq:loop0}
\end{equation}
Hence, ion currents cannot circulate entirely in far field, they must leave and enter regions where $\varrho$ and $g$ vary.  Or, to put it another way, closed paths which `follow the current' in the sense that $\Ivec\cdot\dl>0$ must pass through regions where $\grad\varrho\ne0$ and (possibly separately) where $\grad g\ne0$.  No such paths obtain in the far-field regions where $g$ and $\varrho$ are constant.

\subsubsection{Isolated line sources do not exist}
We can use either of the two results above to conclude that there is no analog of a single, isolated, current-carrying wire in vacuum, in the charge-neutral NP system.  First of all \Eq{eq:loop0} rules it out because in the magnetostatic analogy the integral of the magnetic field around a loop around such a wire would be non-zero, but in our case the integral of our current density (in the far field) must be zero.  Alternatively, \Eq{eq:Sarea0} also rules it out because the existence of an isolated line source would imply the existence of a surface for which $\int\Svec\cdot\dAvec\ne0$, even though $\varrho$ was constant around the perimeter, violating \Eq{eq:oint}.  On the other hand though, the analog of a \emph{pair} of wires carrying equal but opposite currents is allowed, see \Fig{fig:farfield}.  However the net current must always be zero in such a case.  These mathematical statements are equivalent to the fact that it is topologically impossible to isolate a line source $\Svec=\grad g\times\grad\varrho$ without having gradients in $g$ and $\varrho$ persist to infinity (or end at walls).  Also, like the electrical current $\Jvec$ in magnetostatics, since $\div\Svec=0$, a line source of $\Svec$ must either extend to infinity or close in on itself; it cannot start or end in free space.

\begin{figure}[t]
\includegraphics[width=80mm]{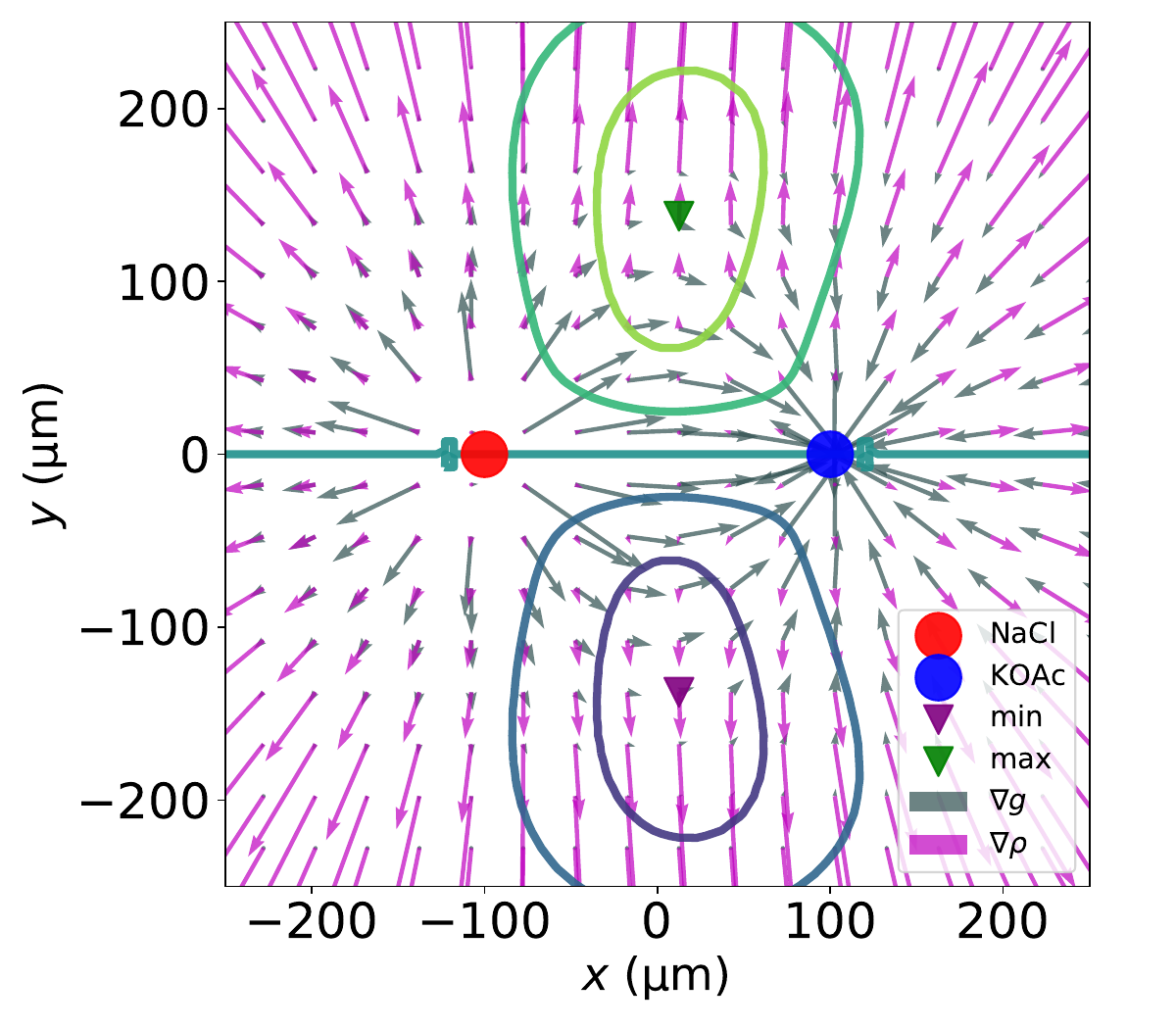}
\caption{The source $S_z$ due to a pair of salt sources in two dimensions \cite{jupyter2}.  The salt sources are taken to be NaCl (red) and KOAc (blue) and are placed $w=\SI{200}{\micro\metre}$ apart along the $x$ axis. The sources have radii $R_s=\SI{25}{\micro\metre}$, and the salts have been diffusing out for $t=4\,\mathrm{s}$.  The source strength $S_z(x,y)$ is shown \via\ contours, with triangles indicating the extrema.  The vector fields $\nabla\rho$ and $\nabla g$ that generate $S_z$ via \Eq{eq:source} are plotted as magenta and grey arrows, respectively. The ratio between the salt concentration in the sources and the background salt is $100:1$.\label{fig:source}}
\end{figure}

\subsection{Autonomous ion currents require two or dimensions and three or more ion species}\label{subsec:sources}
The current-resistivity picture with its emphasis on $\Svec=\grad g\times\grad\varrho$ as a source term explains why three or more ionic species are required to generate an ion current.  Suppose that there are $n$ ionic species and consider the defining relations $g=\sum_i\,z_i D_i c_i$ and $\sigma=\sum_i\,z_i^2D_i c_i$ under the constraints $\sum_i\,z_i c_i = 0$ and $c_i\ge 0$.  The minimum number of ions is $n=2$ since at least two oppositely-charged ion species are needed for charge neutrality.  But if there are $n=2$ ionic species, $g$ and $\sigma$ are not independent quantities; rather, the ion concentrations can be eliminated to find $g=\beta\sigma$ where $\beta=(D_1-D_2)/(z_1D_1-z_2D_2)$ is a proportionality constant (noting that the valencies $z_i$ are signed quantities) \cite{Anderson1989, warren2020}.  It follows that $\grad g\times\grad\varrho=0$ and there can be no ion currents in a binary electrolyte unless ions are injected or removed by Faradaic reactions occurring at the surfaces of electrodes. In the $n=2$ case, since there is no ion current, \Eq{eq:i1} becomes $\grad g=-\sigma\grad\varphi$.  This can be integrated to find $\varphi=\beta\ln\varrho$ to within a constant, which is a well-known result in this context; there are no action-at-a-distance effects, and in the literature one often refers to $\varphi$ as a diffuse liquid junction potential \cite{rieger1994, newman2004}.

Hence $n\ge3$ is required for autonomous ion currents to arise.  In such a situation though, one expects ion currents to be \emph{generically} present unless the gradients of $g$ and $\sigma$ are everywhere parallel.  This situation obviously obtains in purely one-dimensional systems, and therefore two or three dimensions are also required.  Finally, since it is only the two quantities $g$ and $\sigma$ that appear in the theory (\Table{tab:dual}), it follows that systems with $n=4$, 5, \etc\ ion species introduce no new physics not present for $n=3$.

\section{Two dimensional systems}\label{sec:2d}
In many experiments, for example in Williams \etal~\cite{williams2024}, the height is kept small (\eg\ $\lesssim\SI{100}{\micro\metre}$) to avoid convection caused by mass-density gradients coming from the salt concentration gradients \cite{Williams2020, Gu2018, Selva2012}.  The experimental geometry is then largely two dimensional as the lateral ($x$, $y$) dimensions are much larger than the height ($z$).  Motivated by this we now consider two-dimensional geometries.  The prototypical situation is shown in \Fig{fig:source}.  This comprises two (disc-like) sources of diffusing salts of radius $R_s$ with centres $w$ apart, along the $x$-axis.

In two dimensions \Eq{eq:curl2} simplifies considerably since the vector potential is $\Avec=(0,0,A_z)$ and the source is $\Svec=(0,0,S_z)$\,; only the $z$-components are non-zero, and the functions are $A_z(x, y)$ and $S_z(x, y)$.  The current $\Ivec$ is then two-dimensional,
\begin{equation}
    \Ivec=\left(I_x,I_y,0\right)=
    \Bigl(\frac{\partial A_z}{\partial y},
    -\frac{\partial A_z}{\partial x},0\Bigr)\,.\label{eq:2d}
\end{equation}
\Eq{eq:curl2} becomes a pseudo-scalar Poisson-like PDE,
\begin{equation}
  \frac{\partial }{\partial x}
  \Bigl(\varrho\frac{\partial A_z}{\partial x}\Bigr) +
  \frac{\partial }{\partial y}
  \Bigl(\varrho\frac{\partial A_z}{\partial y}\Bigr)+S_z=0\,.
  \label{eq:2da}
\end{equation}
In this form it has the appearance of a steady-state diffusion problem for $A_z(x, y)$ with a position-dependent `diffusion coefficient'~$\varrho$, and a sink term $S_z$. 

In these terms, the boundary condition $\Ivec\cdot\nvec=0$ becomes $n_x{\partial A_z}/{\partial y}-n_y{\partial A_z}/{\partial x}=0$.  If we set $\tvec=\nvec\times\ehat_z$ where $\ehat_z=(0,0,1)$, this can be written as $\tvec\cdot\grad A_z=0$.  This tells us that $A_z$ must be a constant along the insulating boundaries in \Eq{eq:2da}.  If there is only one such boundary we can set the constant to zero without loss of generality.

\subsection{Far field revisited}\label{subsec:ff2d}
We are predominantly interested in studying the non-local currents and fields, so we only consider times such that the salts from the salt sources have diffused a distance of order their separation $w$. Then at distances $\gg w$ (\ie\ in far field), there is only a uniform background where $\varrho$ and $g$ are constant, and $S_z$ vanishes.  In such a region $A_z$ and $\varphi$ are a pair of harmonic functions, because \Eqs{eq:pot} and~\eqref{eq:2da} reduce to  $\delsq\varphi=0$ and $\delsq\!A_z=0$ respectively (incidentally the first of these implies there are no space charges in such a region).  Moreover, it follows from \Eq{eq:2d} and Ohm's law  that
\begin{equation}
  \grad\varphi\cdot\grad A_z
  =\frac{\partial\varphi}{\partial x}\frac{\partial A_z}{\partial x}
  +\frac{\partial\varphi}{\partial y}\frac{\partial A_z}{\partial y}
  =-\frac{I_xI_y}{\raisebox{1pt}{$\varrho$}}
  +\frac{I_yI_x}{\raisebox{1pt}{$\varrho$}}=0\,.
\end{equation}
Hence the gradients $\grad A_z$ and $\grad\varphi$ are orthogonal.  This implies that $A_z$ and $\varphi$ can be matched to the real and imaginary parts of a complex analytic function \cite{mathews1964}, just like the situation encountered for the velocity potential and the stream function in two-dimensional irrotational solenoidal fluid flow \cite{batchelor1967, faber1995}.  In the present case, in order that the Cauchy-Riemann (CR) conditions be satisfied, we should write $w(z)=\varrho A_z+i\varphi$ where $z=x+iy$.  Then Ohm's law implies CR, since
\begin{equation}
  \varrho\Ivec=-\grad\varphi \>\implies\>
    \varrho\frac{\partial A_z}{\partial x}=
    \frac{\partial\varphi}{\partial y}\,,\quad
    \varrho\frac{\partial A_z}{\partial y}=
    -\frac{\partial\varphi}{\partial x}\,.
\end{equation}
Any such analytic function $w(z)$ will provide a solution of the governing equations, but only in regions where the conditions stated at the start hold (constancy of $\varrho$ and $g$).

As we are far ($\gg w$) from the source, we can use the far-field solutions of the two-dimensional Laplace equation.  As is well known such solutions can be represented by a multipole expansion \cite{batchelor1967, jackson1999, joslin1983}.  Working in polar co-ordinates and focusing on the vector potential $A_z(r,\theta)$, we can write
\begin{equation}
  \varrho A_z=a_0\ln\frac{1}{r}
  +\sum_{n=1}^\infty\frac{a_n\cos n\theta
    +b_n\sin n\theta}{r^n}\,,\label{eq:multipole}
\end{equation}
where an overall additive constant has been omitted since it does not affect the physics. 

Just as in the three-dimensional case, the amplitude $a_0$ of the leading (monopole) term here must be zero. To show this we start by rewriting \Eq{eq:2da} as
\begin{equation}
\div(\varrho\grad A_z)+S_z=0\,.
  \label{eq:2da2}
\end{equation}
Then we integrate this equation over an area $A$ and exploit the divergence theorem for the first term
\begin{equation}
  {\textstyle\int}\varrho \grad A_z\cdot\nvec\,\ds
  +{\textstyle\int} S_z\,\dA=0\,,
  \label{eq:2dAint}
\end{equation}
where the first integral is over the perimeter with $\nvec$ the unit normal to this perimeter.

Now we specialise to a disc $\cal D$ with a perimeter entirely in the far field. Now from \Eq{eq:Sarea0} the second integral in \Eq{eq:2dAint} is zero, and $\varrho$ is a constant on the perimeter, so we have
\begin{equation}
  \varrho{\textstyle\int}\grad A_z\cdot\nvec\,\ds=0
\end{equation}
(note this does \emph{not} require $\varrho$ is constant in the interior).   Injecting the multipole solution in \Eq{eq:multipole},
%
the monopole term (only) contributes to the integral, and so $a_0$ must be zero and the leading order term is the dipole term.

\begin{figure}[t]
\includegraphics[width=50mm]{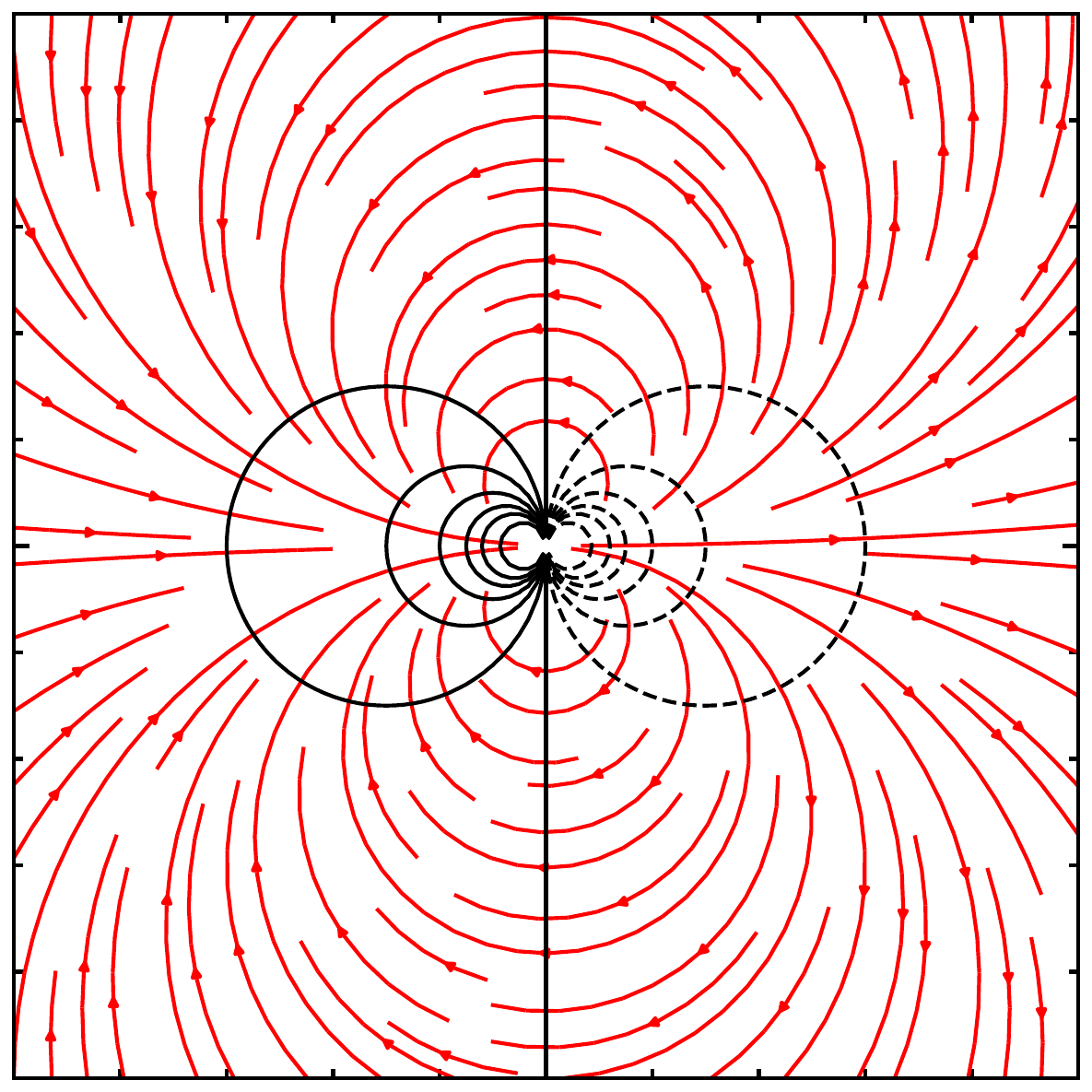}
\caption{Ion current (red lines with arrows) and equipotential lines (black lines) around a dipole line source localised at the origin, oriented to mimic the salt sources in \Fig{fig:source}.  This is the two-dimensional analog of \Fig{fig:farfield3d}.  The length scale is arbitrary.\label{fig:farfield}}
\end{figure}

We now consider this dipole term in \Eq{eq:multipole}.  In terms of the complex potential introduced above, this term corresponds to $w(z)=c_1/z$, where $c_1=a_1+ib_1$.  From this, the dominant contribution in far field, in two dimensions, is
\begin{equation}
  \varrho A_z=\frac{a_1\cos\theta+b_1\sin\theta}{r}\,,\quad
  \varphi=\frac{b_1\cos\theta-a_1\sin\theta}{r}\,.\label{eq:dipole1}
\end{equation}
The corresponding current (\eg\ from $\varrho\Ivec=-\grad\varphi$) is 
\begin{equation}
  I_x=\frac{a_1\sin2\theta-b_1\cos2\theta}{\varrho r^2}\,,\quad
  I_y=-\frac{a_1\cos2\theta+b_1\sin2\theta}{\varrho r^2}\,.\label{eq:dipole2}
\end{equation}
An example of this far-field dipolar solution is shown in \Fig{fig:farfield} for $a=0$ and $b=-1$, to match the orientation of the sources in \Fig{fig:source}.  \Eq{eq:dipole2} means that generically, in two dimensions, the electric field and current in far field obey
\begin{equation}
  |\Evec|=\varrho|\Ivec|={|c_1|}/{r^2}\label{eq:dipole3}
\end{equation}
around a localised source, characterised by $c_1=a_1+i b_1$.

As in three dimensions, this amplitude can only be estimated, since the essential presence of crossed gradients in \Eq{eq:current} implies that $\varrho$ must be spatially varying in the vicinity of the source\,; \confer\ argument below \Eq{eq:oint}.  
This precludes an exact solution of \Eq{eq:2da}.  Nevertheless, let us proceed here by introducing a `decoupling' approximation in which we assume $\varrho$ is constant, even in the source region where $\Svec=\grad g\times\grad\varrho\ne0$.  With this, \Eq{eq:2da} reduces to a Poisson equation, $\delsq(\varrho A_z)+S_z\simeq0$ which has a formal solution in terms of a two-dimensional Green's function,
\begin{equation}
  \varrho A_z(\rvec) = -({2\pi})^{-1}{\textstyle\int}\ln|\rvec-\rvec'|\,
  S_z(\rvec')\,\dA'.
\end{equation}
If we expand assuming that $S_z$ is localised to a region around the origin, we recover the far-field solution as in \Eq{eq:multipole} with
\begin{subequations}
  \begin{align}
    &a_0=(2\pi)^{-1}{\textstyle\int} S_z\,\dA=0\,,\\[3pt]
    &a_1\simeq(2\pi)^{-1}{\textstyle\int}x\,S_z\,\dA\,,\quad
    b_1\simeq(2\pi)^{-1}{\textstyle\int}y\,S_z\,\dA\,.\label{eq:abcoeffs}
  \end{align}
\end{subequations}
The result for $a_0$ is validated by the exact result above, and those for $a_1$ and $b_1$ can be used to estimate the leading-order behaviour in far field in terms of the moments of $S_z$.

\begin{figure}[t]
\includegraphics[width=80mm]{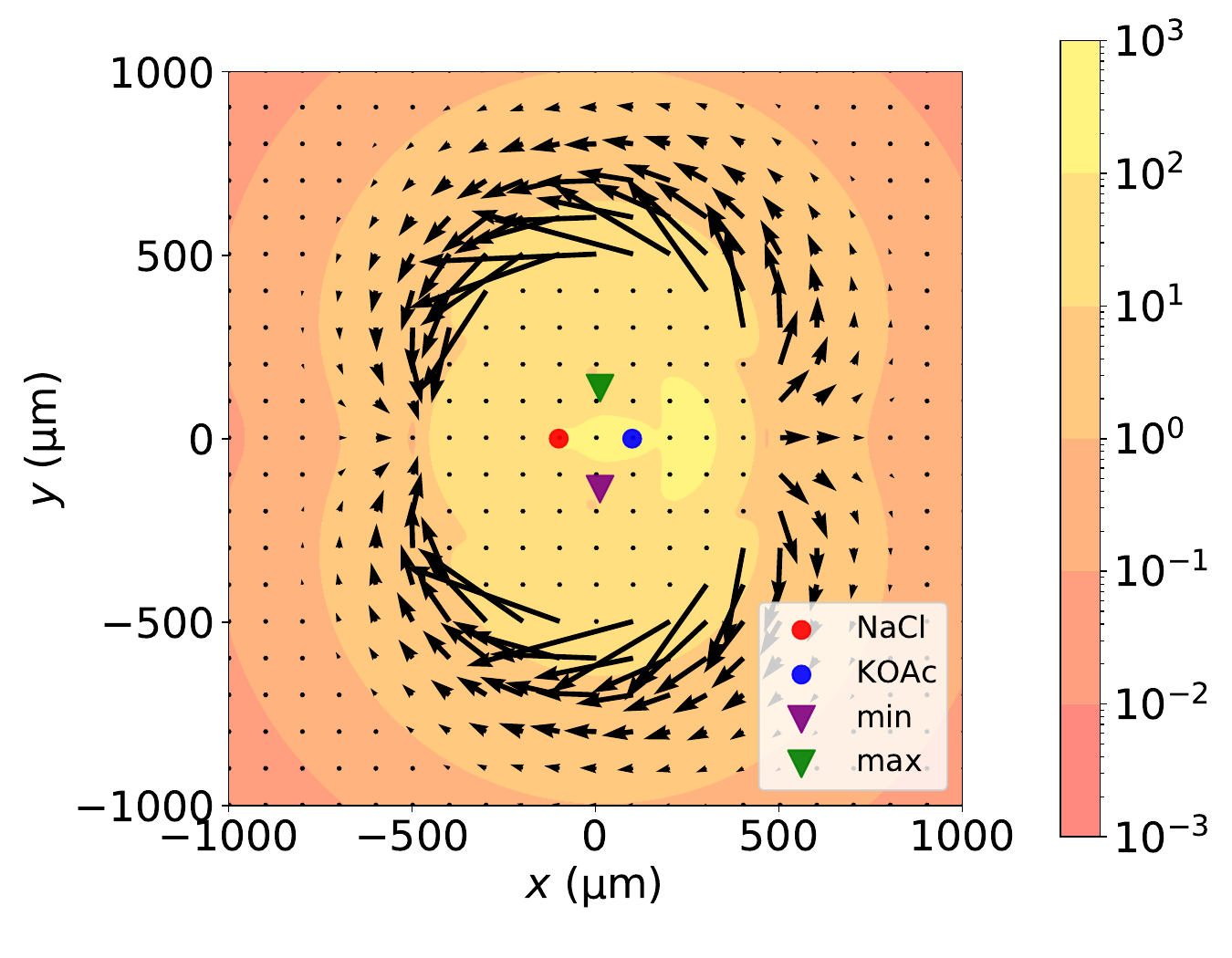}
\caption{Plot of the electric field $\Evec$ as a function of position, in units of $\mathrm{V}\,\mathrm{m}^{-1}$.  The arrows show $\Evec$ as a vector field while the shading shows the magnitude of $\Evec$. For clarity, arrows are not shown in a central disc of radius $500\,\um$.  The geometry and salt sources are as in \Fig{fig:source}, and we show the positions of the salt sources and maximum and minimum in $S_z$.  The pattern in far field matches that in \Fig{fig:farfield}.\label{fig:E}}
\end{figure}

It is worth noting that the equations we use in far-field region are not only analagous to equations for magnetostatics in a vacuum near current-carrying wires \cite{jackson1999, feynman1964}, but also analogous to the equations for fluid flow near a localised source of vorticity \cite{batchelor1967}.  Examples of this are vortex lines produced by aircraft wings at their tips, and tornadoes.

\subsection{Practical estimates of the size of the effects}
Having obtained expressions for the electric field, we want to use them to obtain estimates of its strength. We consider a simple two-dimensional system of a pair of salt sources at a distance $w$ apart at a time such that the salts from them have diffused of order the same distance, $w$. Then the source dipole has only one lengthscale, $w$.  We consider the far-field situation, where we can use \Eqs{eq:dipole3} and~\eqref{eq:abcoeffs} for $\Evec$ and $c_1=a_1+ib_1$. In our reduced units, the coefficients have units of length and so scale as $w$, so in these units $|\Evec|\sim w/r^2$.  As we have already noted in the context of \Eq{eq:eff}, electric fields in our system have units of inverse length.  Restoring the full units, we therefore expect in two dimensions the electric field in the far field of a localised source to be dipolar with
\begin{equation}
    |\Evec|\sim ({\kT}/{e})\times w/r^2\qquad(r\gg w)\,.\label{eq:escale}
\end{equation}
This is the two-dimensional analog to \Eq{eq:eff}.  For example, at a distance $r\simeq100\,\um$ from a pair of sources $w\simeq10\,\um$ apart, the electric field is of the order $10\,\mathrm{V}\,\mathrm{m}^{-1}$.  This is around one tenth of the electric field strength estimated by McDermott \etal~\cite{mcdermott2012} to be present $100\,\um$ away from a dissolving $10\,\um$ calcium carbonate crystal, due to the diffuse liquid junction potential arising from the \ce{CaCO3} gradient. 

\Fig{fig:E} shows the electric field around the pair of salt sources shown in \Fig{fig:source}, obtained by solving the governing equations numerically. For details see Appendix \ref{app:numerics} \cite{jupyter2}.  As predicted, in the far-field region the electric field is at most of order $10\,\mathrm{V}\,\mathrm{m}^{-1}$.  Note also the characteristic dipolar nature of field.  Near the salt sources where gradients are present, the electric field is an order of magnitude higher, several $100\,\mathrm{V}\,\mathrm{m}^{-1}$.  

Electric fields of the magnitude given by \Eq{eq:escale} imply a potential difference over a distance order $r$ of order $\Delta V\sim (\kT/e)(w/r)$.  For example at $r\sim 10\,w$ distant from a pair of salt sources, $\Delta V$ is of order a few mV. The current depends on the conductivity $\sigma$ which is proportional to salt concentration. For salt concentrations of order 100~mM, $\sigma$ is of order $\SI{1}{\siemens\per\metre}$, and electric fields of order $10\,\mathrm{V}\,\mathrm{m}^{-1}$ give us current densities of order tens of $\SI{}{\ampere\per\metre\squared}$.

\begin{figure}
{\large (a)}\includegraphics[width=80mm]{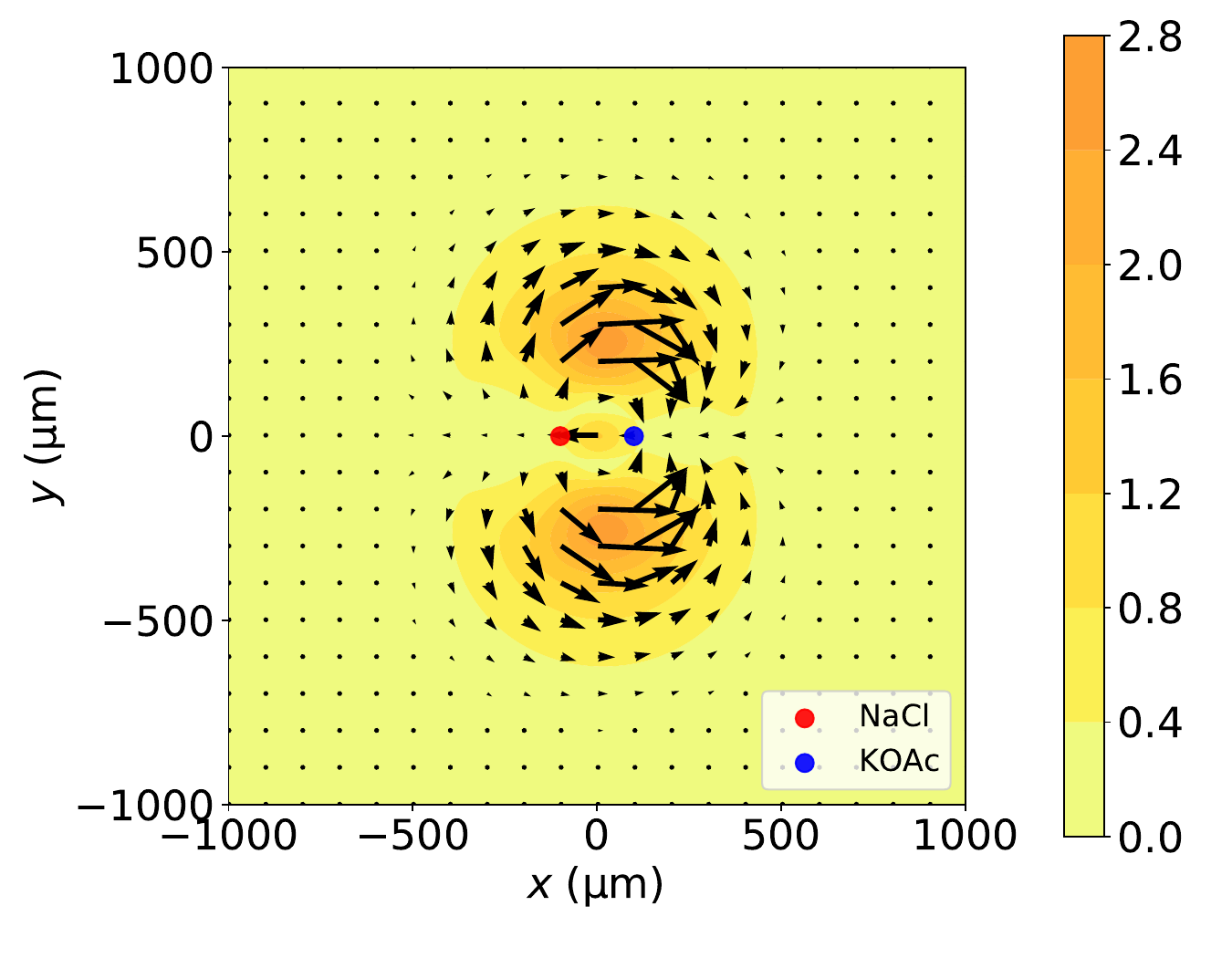}\\
{\large (b)}\includegraphics[width=80mm]{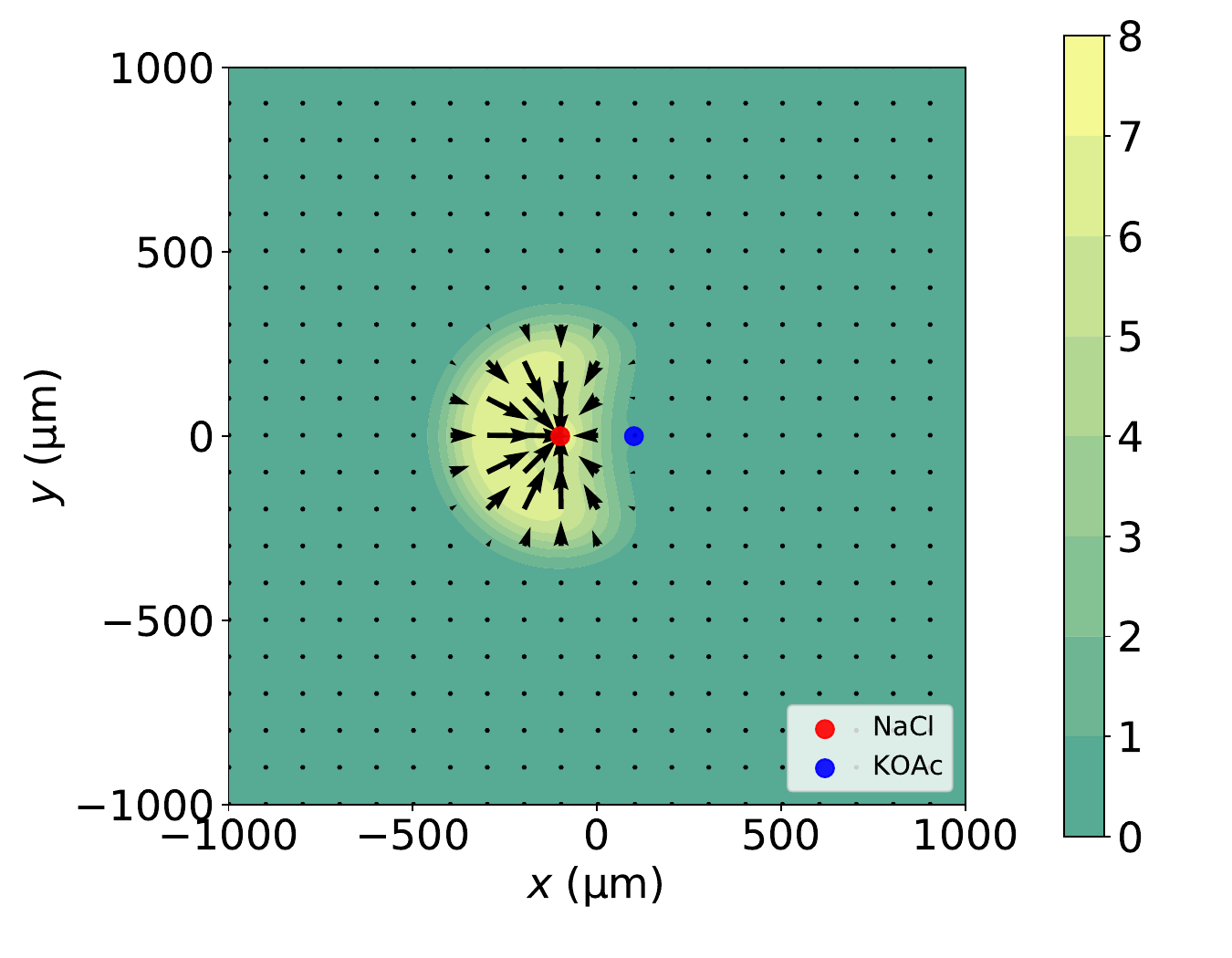}\\
{\large (c)}\includegraphics[width=80mm]{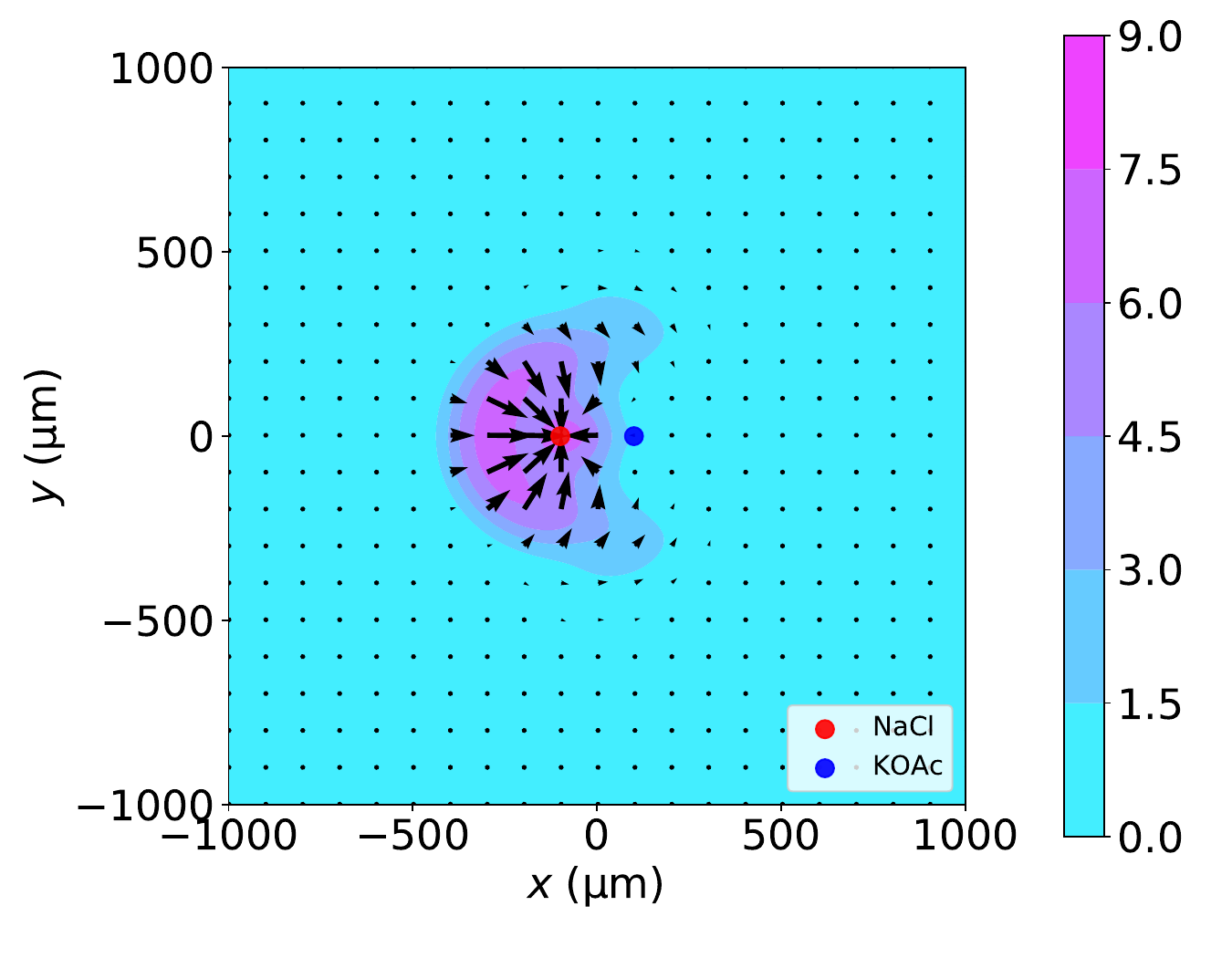}
\caption{\label{fig:saltpair_DP} Examples of the diffusiophoretic (DP) drift velocity of a colloid particle near a pair of salt sources: (a) DP velocity due to the non-local term only, (b) local DP velocity, and (c) combined total DP velocity.  Salt sources and electric field as in \Figs{fig:source} and~\ref{fig:E}. In all three plots the arrows show the velocity while the color scale shows the speed in $\um\,\persec$. The $\zeta$-potential of the colloid is $-50\,\mathrm{mV}$.}
\end{figure}

\section{Diffusiophoresis}

The salt gradients and resulting electric fields will move particles and macromolecules in solution, if they have non-zero zeta ($\zeta$) potentials, as most do. This motion due to a gradient in a solute is called diffusiophoresis (DP).

\subsection{Diffusiophoresis in the far field}

In the far field there are no salt gradients, so any local gradient-driven DP effect vanishes, leaving only electrophoresis in the non-local electric field.  In this region, the DP drift velocity $\Uvec$ is given by the standard electrophoretic expression of Smoluchowski \cite{newman2004}
\begin{equation}
\Uvec=({\epsilon_r\epsilon_0\zeta}/{\eta})\,\Evec\,,
\end{equation}
where $\epsilon_r$ is the relative dielectric permittivity of the medium, $\epsilon_0$ is the permittivity of free space, $\zeta$ is the electrophoretic surface potential of the particle, and $\eta$ is the viscosity.  Using \Eq{eq:escale}, we estimate
\begin{equation}
    |\Uvec|\sim \frac{\epsilon_r\epsilon_0\zeta}{\eta}\,\frac{kT}{e}\times \frac{w}{r^2}\qquad(r\gg w)\,.
    \label{eq:uDPscale}
\end{equation}
In the far-field region, the characteristic DP speed has the same $r^{-2}$ scaling as both the electric field and ion current.  

Note that in the far field $\div\Ivec=0$ implies $\div\Uvec=0$.  This has the consequence that DP in far field cannot cause particles to accumulate. 

Colloidal particles are typically negatively charged with a $\zeta$-potential of order tens of mV. For example, Williams \etal~\cite{williams2024} worked with colloids with $\zeta=-\SI{50}{\milli\volt}$. For such particles the electrophoretic mobility $\epsilon_r\epsilon_0\zeta/\eta\sim 10^{-8}\,\mathrm{m}^2\,\mathrm{s}^{-1}\,\mathrm{V}^{-1}$ in water, assuming a viscosity $\eta\simeq1\,\mathrm{mPa}\,\mathrm{s}$ and a relative permittivity $\epsilon_r\simeq80$.  With a far-field electric field of the order $10\,\mathrm{V}\,\mathrm{m}^{-1}$, this gives DP speeds of order $0.1\,\um\,\persec$.

\subsection{Local and non-local diffusiophoresis}

The full expression for the diffusiophoretic drift velocity $\Uvec$ is  
(see \Appendix{app:3DP} and Warren \cite{warren2020} for details)
\begin{equation}
  \Uvec/\Gamma_0=\alpha \grad\ln c+\gamma\varrho\grad g+\gamma\varrho\Ivec\,.\label{eq:3DP}
\end{equation}
The first term on the right-hand side is chemiphoresis in the total ion concentration gradient $c=\sum_ic_i$ (for simplicity we assume univalent ions). The second term is local electrophoresis due to the local gradient in $g$, while the final term is non-local electrophoresis driven by the electric field associated with the ion current.

The prefactor $\Gamma_0=(\epsilon_r\epsilon_0/\eta)(\kT/e)^2 \simeq 510\,\um^2\,\persec$ in water at room temperature.  The dimensionless coefficients $\alpha=4\ln\cosh(e\zeta/4\kT)$ and $\gamma=e\zeta/\kT$ depend on the particle $\zeta$-potential and are of order unity.  Note that the three contributions generally point in different directions and have to be added vectorially \cite{warren2020}.  

For a binary electrolyte $g=\beta\sigma$ as explained in section \ref{subsec:sources}, and $\sigma\propto c$, so that with $\varrho=1/\sigma$ the first two contributions can be combined into an overall term proportional to $\grad\ln c$.  Also, the ion current vanishes so that the third term does not appear.  So in this case (binary electrolyte), the DP velocity can be written as $\Uvec=\Gamma_0(\alpha+\beta\gamma)\grad\ln c$.  This is the usual well-known result found in the literature \cite{Anderson1989}.

Continuing with our example of a pair of salt sources, we calculated the contributions to the DP velocity in \Eq{eq:3DP}, and show the results in \Fig{fig:saltpair_DP}.  As we would expect from our magnetostatic analogy, the non-local DP velocity field resembles the magnetic field lines around a pair of wires carrying equal and opposite currents, \confer\ \Fig{fig:farfield}. There are two counter-rotating vortices centered on the extrema of the source term (see \Fig{fig:source}), \emph{not} on the salt sources themselves.  The far field of the non-local DP velocity has the same pattern as the electric field in \Fig{fig:E}, as it must.

If we compare the plots of the non-local DP velocity (\partFig{fig:saltpair_DP}{a}) with the local DP velocity (\partFig{fig:saltpair_DP}{b}), we see that the local DP speeds peak at around ten times the non-local DP speed, with a magnitude of order a few $\um\,\persec$ as opposed to a few tenths of $\um\,\persec$, but the non-local DP velocity is much longer ranged. The salt gradients have only spread a distance of at most $100\,\um$ away from the sources, and so local DP is almost zero outside this range since the diffusion profiles decay as Gaussians in far field.  However, non-local DP occurs throughout our system, which spans $2000\,\um$.  This indicates the long-ranged, non-local character of the phenomenon.

In a system confined by plates in the $z$-direction, one may also have to account for diffusio-osmosis (DO) occurring at the walls.  This can drive fluid motion in the gap, which advects the particles and should be compounded with DP phenomena discussed above.  In the experimental studies of Williams \etal~\cite{williams2024}, there was evidence that this effect was not very large, so it is not necessarily the case that DO makes a significant contribution.  For completeness, we discuss the effects of DO in \Appendix{app:do}.

\section{Conclusion}
Using analogies with textbook magnetostatics, we have shown that a pair of different salt sources a distance $w$ apart generates a dipolar electric field which decays with distance $r$ from the sources as $(\kT/e)\times w^2/r^3$ in three dimensions (section \ref{subsec:ff3d}) or as $(\kT/e)\times w/r^2$ in two dimensions (section \ref{subsec:ff2d}).  This field  is typically weaker than the local electric fields generated by local salt gradients. However, it propagates almost instantaneously, \ie\ on the Debye time (see \Appendix{app:neut}).  As this time is so short, in practice the relevant time scale for the electric field to appear is set by the time it takes for the salt gradients spreading out by diffusion to meet and start to cross.  This diffusion time scale is of order $w^2/D$.  

As propagation is almost instantaneous, the whole solution senses any change in the pair of salt sources effectively instantly, with a signal strength that decays as a power law $r^{-d}$ in $d$ dimensions (\ie\ not exponentially, like the individual salt gradients).  This presents an interesting opportunity in the context of intracellular signalling.  In particular the limit identified by Bryant and Machta \cite{bryant2023} for the speed of transmission of information \via\ diffusion does not apply if the source of the information comes from a pair of sources of different salts.  Then the rate of change of the signal is always set by the diffusion time across the source, \ie\ $w^2/D$, regardless of the distance of transmission.  The signal arising from this (\ie\ the ion current and associated electric field) is effectively broadcast almost instantaneously across the whole cell, or between one cell and a neighbour.

The electric potential / ion current signal propagates even in the presence of large background salt concentrations, such as the order 100\,mM concentrations in living cells. The electric potential changes may be small, of order a few mV, but ion channels can detect electrical potential changes of this size \cite{klejchova2021, lazzari2021}. So it is possible that the non-local fields studied here could have been adapted for signalling in some cells, although we know of no evidence for this.  Even if these non-local fields are not used for signalling, then the large fluxes of multiple charged species (sodium, potassium, chloride, ATP, \etc) means the non-local fields are inevitable in living cells, and may need to be filtered out by cell signalling processes that do rely on detecting subtle changes in electrostatic potentials.

The emerging field of iontronics \cite{han2022,bocquet2023} may also be able to exploit the non-local effects studied here. Iontronics, simply speaking, uses ion fluxes (in solution) to perform computations, rather than the electrical currents that used in electronic devices such as silicon chips \cite{han2022, bocquet2023}. The non-local effect studied here may make it possible for one iontronic element to pass information on to multiple distant iontronic elements effectively instantaneously.

To observe the effects of the non-local field in experiment, we considered a pair of salt sources. These can be created in a number of ways. In the work of Williams \etal~\cite{williams2024}, the salt sources were the soluto-inertial gel beacons of Banerjee and coworkers \cite{Banerjee2016}.  However this is not the only way to create crossed salt gradients. McDermott \etal~\cite{mcdermott2012} showed that dissolving crystals create salt concentration gradients around them. In their case, they used calcium carbonate crystals. So pairs of different dissolving salt crystals will also work.  Salts can also be simply pumped into the solution via small pipettes, see for example the work of Secchi and coworkers \cite{Secchi2016,secchi2017} or Katzmeier and Simmel \cite{katzmeier2024}.  So a pair of pipettes pumping out different salt solutions would be a third option.  Note that the flow rates can accelerate the formation of the crossed gradients, but that the propagation of any flow pattern is orders of magnitude slower than propagation of the electric field.

Another way to create crossed gradients is in a microfluidics device in which streams of different salts can be merged \cite{Shin2020}.  This offers up the interesting prospect of a device which can maintain \emph{continuous} crossed salt gradients in steady-state.  Note that advection \perse\ does not affect the formation of ion currents in the NP equations \cite{probstein1994}, since it couples to the space charge which is vanishingly small.  On the other hand the ion currents, if present, would perturb the concentration profiles of the salt streams in a way that might be measurable.

So there are a number of ways to create the large crossed salt gradients that maximise the currents and electric fields. However, so long as there are (not-parallel) gradients in three or more ionic species, the currents and fields will be present. Even in pure water there are always \ce{H+} and \ce{OH-} ions, in addition to the ions of any salt(s) added; and the number of species increases in any solution exposed to the air since atmospheric carbon dioxide can dissolve in the solution, creating pH gradients \cite{shin2017} --- which are ionic gradients. So even for simple solutions of a single salt, although the effects studied here may be very weak, there may in practice be gradients of three or more ions and so ion currents will be present.

Last, our work can also be used as an interesting case study in vector calculus in a setting which differs from the usual ones of electromagnetism and fluid dynamics in certain crucial ways, and as such it may be of pedagogical interest.  One example is the generalised Biot-Savart law in \Appendix{app:BS}.  There are still a considerable number of open problems in the space we have described, for example the determination of the boundary conditions across gradients of $\grad g$ and $\grad\varrho$ which are sharp enough to be regarded as being discontinuous jumps.  One could then consider the intersection of two such jumps  as corresponding to an idealised line source as in \Fig{fig:crossgrad}, with some hope of obtaining an exact piecewise-continuous solution for $\varphi$ which according to \Eq{eq:pot} should satisfy a simple Poisson equation $\delsq\varphi=0$ in each sector.  Since the integral in the generalised BS law in \Eq{eq:genBS} is easy to evaluate for a line source, one could then assess the role of the gradient term $\grad\psi$.  Another example could be the intersection of two spherical regions providing an idealised circular line source as a model for the situation in \Fig{fig:farfield3d}.  

{\bf Acknowledgements} --- It is a pleasure to acknowledge discussions with I.~Williams and J.~Keddie. Funding was provided by the EPSRC through a New Horizons grant (Grant No. EP/V048473/1).

\appendix

\section{Charge neutrality and time scales in the Nernst-Planck equations}\label{app:neut}
We return to the question of charge neutrality in the NP equations, which is closely connected to the consideration of the time scales on which the ion currents and associated electric fields develop.  The material here has been summarised in \Refcite{warren2020} and presented in more detail in Williams \etal~\cite{williams2024}\,; further relevant background can be found in Bazant \etal~\cite{bazant2004}.

Our starting point is to estimate the space charge density from the Poisson equation, in the form \cite{jackson1999}
\begin{equation}
  \epsilon_r\epsilon_0\delsq\phi=-e\textsumi\,z_i c_i\,.
\end{equation}
Note that it is $\phi=\varphi\kT/e$ that features here.  Let us introduce now the Debye length $\lD$ defined by \cite{newman2004}
\begin{equation}
  \lD^{-2}=\frac{e^2\,\textsumi\,z_i^2c_i}{\epsilon_r\epsilon_0\kT}\,,
\end{equation}
where $\sum_iz_i^2c_i$ is a quantity known as the ionic strength.  Using the Debye length, the Poisson equation rewrites as
\begin{equation}
  \lD^2\delsq\varphi
  =-(\textsumi\,z_ic_i)/(\textsumi\,z_i^2c_i)\,.\label{eq:app:poisson}
\end{equation}
The left-hand side is now in terms of the dimensionless electrostatic potential $\varphi$, which is of order unity in the NP problem.

In \Eq{eq:app:poisson} we see that large fractional deviations from charge neutrality (\eg\ of order 10\%) cause the potential $\varphi$ to vary over a length scale of the Debye length. This is what happens in electrtical double layers (EDLs). So for systems $L\gg\lD$ across, the Poisson equation is effectively singular in the sense that at boundaries with imposed potentials, charge neutrality is violated in the EDLs where the potential varies rapidly (over $\lD$), but outside of these EDLs the solution is very close to charge neutrality, with $\sum_iz_ic_i\simeq0$.  This is the essential justification for the charge neutrality assumption in the NP problem.

We can now estimate the timescale for the charge movements needed to obtain steady electric fields, in a cubic system with sides $L$.  The timescale $\tau\sim Q_\text{S}/J$: the total space charge charge $Q_\text{S}$ divided by the current $J$ \cite{horowitz2015}. The total charge density $\sim c_s$ for $c_s$ the salt concentration, and from the Poisson \Eq{eq:app:poisson} the fractional charge density needed for the potential to vary over a lengthscale $L$ is $(\lD/L)^2$. So $Q_\text{S}\sim(\lD/L)^2c_sL^3\sim\lD^2Lc_s$.  The current density $I=\sigma E$, with conductivity $\sigma\sim Dc_s$, where $D$ is a typical ion diffusion coefficient. The current $J=IL^2$ and with $E\sim 1/L$ in units of $\kT/e$ per unit length, we have $J\sim Dc_s(1/L)L^2\sim Dc_sL$.  Therefore the time scale is
\begin{equation}
  \tau\sim \lD^2/D \,.
\end{equation}
This is often called the Debye time \cite{bazant2004, rubinstein2009, barnaveli2024}.

\begin{table}
\begin{ruledtabular}
  \begin{tabular}{rccccc}
    ionic strength & $\lD$ & $\lD^2/L^2$ & $\lD^2/D$ & $\lD L/D$ & $L^2/D$ \\[2pt]
    \hline\\[-7pt]
    1\,mM   & 10\,nm & $10^{-8}$  & 0.1\,$\upmu$s & 1\,ms   & 10\,s \\
    100\,mM & 1\,nm  & $10^{-10}$ & 1\,ns         & 0.1\,ms & 10\,s
\end{tabular}
\end{ruledtabular}
\caption{Examples of relaxation times in an electrolyte, calculated for $D=10^{-9}\,\mathrm{m}^2\,\mathrm{s}^{-1}$ and $L=100\,\um$.\label{tab:tscales}}
\end{table}

As Bazant \etal~\cite{bazant2004} explain, there is a \emph{further} time scale of order $\lD L/D$ associated with EDL charging.  This result can be obtained by adapting the above RC circuit analogy to the EDL case \cite{warren2023}.  The total charge in the EDLs can be estimated by multiplying the ionic strength by the surface area $\sim L^2$ and the EDL thickness $\sim\lD$.  This gives $Q_\text{EDL}\sim \lD L^2$. This charge is larger than $Q_\text{S}$  by a factor of $L/\lD$ and requires the longer timescale to be reached in order to relax.

To summarise, our central assumption is that $L\gg\lD$.  This not only justifies the assumption of local charge neutrality in the bulk, but also enforces a strict time scale separation and hierarchy $\lD^2/D\ll \lD L/D\ll L^2/D$.  This means that the analysis in the main text is applicable whether or not the EDLs have fully relaxed.  \Table{tab:tscales} shows the typical values of these time scales, calculated for two different ionic strengths.  It is apparent that the time scale separation and assumption of local charge neutrality are well justified in these problems.

\section{Generalised Biot-Savart law for systems with crossed salt gradients}\label{app:BS}
An obvious question to ask is what is the analog in the NP equations of the Biot-Savart (BS) law in magnetostatics \cite{jackson1999},
\begin{equation}
  \Bvec=\frac{\mu_0}{4\pi}\int
  \frac{\Jvec(\rvec')\times(\rvec-\rvec')}{|\rvec-\rvec'|^3}\,\dV'\,.
\end{equation}
Usually in electromagnetism, BS is taken as a postulate and the differential forms of Amp\`ere's and Gauss' laws given in \Table{tab:mag} are deduced from this \cite{jackson1999, griffiths2013}.  Conversely, one can deduce the BS law from  Amp\`ere's and Gauss' laws using the Helmholtz theorem \cite{griffiths2013}.  Unfortunately in the NP case, the troublesome $\varrho$ gets in the way.  Let us start by providing the correct analog of the BS law in the NP case,
\begin{equation}
  \varrho\Ivec=\frac{1}{4\pi}\int
  \frac{\Svec(\rvec')\times(\rvec-\rvec')}{|\rvec-\rvec'|^3}\,\dV'
  +\grad\psi\,.\label{eq:genBS}
\end{equation}
The critical difference here is the additional presence of the gradient of a scalar quantity $\psi$.  This is essential to allow $\div(\varrho\Ivec)\ne0$, which would otherwise follow as a vector calculus identity from the first term (see next).  The scalar field is determined by the requirement that $\div\Ivec=0$.  As a corollary to the uniqueness result discussed in section \ref{sec:ion_current}, this determines $\psi$ to within an additive constant.

To see what this involves, let us start by defining $\Rvec$ as the integral in \Eq{eq:genBS} above.  With this, one can show \cite{jackson1999, griffiths2013}
\begin{equation}
  \curl\Rvec=\Svec\,,\quad\div\Rvec=0\,.\label{eq:I}
\end{equation}
The first expression here is what we want since it reproduces \Eq{eq:current}.  The second however presents a problem since as mentioned in general $\div(\varrho\Ivec)\ne0$.  It is to accommodate this that $\grad\psi$ is introduced in \Eq{eq:genBS} since it does not affect the first of \Eqs{eq:I}.  With this in place, and recalling $\sigma=1/\varrho$, \Eq{eq:genBS} can be recast as
\begin{equation}
  \Ivec=\sigma\Rvec+\sigma\grad\psi\,.
\end{equation}
Hence $\psi$ should be chosen to satisfy
\begin{equation}
  \div(\sigma\grad\psi)+\div(\sigma\Rvec)=0\,.\label{eq:N}
\end{equation}
This resembles \Eq{eq:pot}\,; to solve it we need to evaluate the source term $\div(\sigma\Rvec)$.  In principle this can be done since the integral $\Rvec$ is in terms of known quantities.  Another approach derives from the fact that $\Svec=\curl(-\varrho\grad g)$.  This means $\curl(\Rvec + \varrho\grad g) = 0$ from the first of \Eqs{eq:I}, and hence
\begin{equation}
  \Rvec=-\varrho\grad g-\grad\chi\,,\label{eq:R}
\end{equation}
where we have introduced a further scalar (with a negative sign for convenience) to offset the fact that in general $\div(\varrho\grad g)\ne0$.  Since $\div\Rvec=0$ the scalar quantity $\chi$ should solve
\begin{equation}
  \delsq\chi+\div(\varrho\grad g)=0\,.\label{eq:C}
\end{equation}
We deduce from \Eq{eq:R} that
\begin{equation}
  \div(\sigma\grad\chi)+\delsq g
  +\div(\sigma\Rvec)=0\,.\label{eq:S}
\end{equation}
This gives us the source term in \Eq{eq:N}.  Combining \Eqs{eq:N} and~\eqref{eq:S} we see that in fact $\chi-\psi$ satisfies \Eq{eq:pot}. Conversely, if we solve \Eq{eq:pot} for $\varphi$ and \Eq{eq:C} for $\chi$, we can calculate $\psi=\chi-\varphi$ to use in the generalised BS law in \Eq{eq:genBS}.  But this overlooks the fact that if we do have access to $\varphi$, it is trivial to calculate $\varrho\Ivec$ from \Eq{eq:i1}.  In fact, $\Rvec$ can be eliminated between \Eqs{eq:genBS} and~\eqref{eq:R} to find
$\varrho\Ivec={}-\varrho\grad g-\grad\chi+\grad\psi$.  With the aid of $\varphi=\chi-\psi$ this indeed simplifies to \Eq{eq:i1}.

Solving \Eq{eq:N} to obtain $\psi$ may therefore be feasible but involves just as much work as solving \Eq{eq:pot} for the electrostatic potential $\varphi$, from which $\varrho\Ivec$ follows straightforwardly.  Hence, as a practical means to calculate the ion current, the generalised BS law does not seems to provide much benefit.

The generalised BS law and the complications that ensue provide an interesting pedagogical counterexample to the simpler BS law in electromagnetism.  In that latter case, coming from Amp\`ere's and Gauss' laws, the corresponding $\grad\psi$ is formally still present but can be shown to vanish as a consequence of the simpler structure of the governing equations \cite{griffiths2013}.

\section{Methodology for numerics in two dimensions}\label{app:numerics}
Our results in two dimensions, section \ref{sec:2d}, were obtained by the Python Jupyter notebook included in the supplementary material.  We discretise the relevant PDEs on a two-dimensional square grid, for a region of interest 4\,mm a side, see for example \Fig{fig:E}. We outline how the calculations are done in this appendix, for further details see the Jupyter notebook \cite{jupyter2}.

Note that the theory we develop here relies on a separation of timescales between the rate of evolution of the salt gradients (and hence $g$, $\varrho$ etc), which is slow, and evolution of electric field $\Evec$ and currents $\Ivec$, which are much faster. So the exact way to solve our equations is to start with initial salt gradients, and boundary conditions. Then to use the salt gradients and boundary conditions to solve for the potential and currents, and use these and the gradients to integrate forward the salt profiles to the next time step. By repeating these steps, the system can be evolved forward in time.

\begin{table}
\begin{ruledtabular}
  \begin{tabular}{llllc}
    salt & $D_1$ (cation) & $D_2$ (anion) & $D_s$ (salt) & $\beta$\\[2pt]
    \hline\\[-7pt]
    NaCl & 1.33 & 2.03 & 1.61 & $-$0.21\\
    KOAc & 1.96 & 1.09 & 1.40 & $+$0.29
\end{tabular}
\end{ruledtabular}
\caption{Diffusion coefficients of the ions used in the calculations in section \ref{sec:2d}, from Williams \etal~\cite{williams2024}, in units of $10^{-9}\,\mathrm{m}^2\,\mathrm{s}^{-1}$.  The third and fourth columns refer to the salt diffusion coefficient, and the dimensionless diffusivity contrast that appears in section \ref{subsec:sources}.\label{tab:ions}}
\end{table}

We need the salt gradients at one instant in time, to produce illustrative fields, currents and DP velocities. To do this we do not solve our equations exactly, we do as follows. 
\begin{enumerate}
\item Our salt sources are a pair of discs of radius $R_s=\SI{25}{\micro\metre}$ at constant high salt concentration $c_s$, that model the soluto-inertial beacons of Banerjee \etal~\cite{Banerjee2016}. These sources are centred at  $x=\pm w/2$ and $y=0$. To produce salt profiles, we start with the salt concentration equal to a background concentration $c_b$ everywhere outside the beacons. Then the model profiles at time $t$ are obtained by assuming each salt diffuses independently out from its fixed concentration source, with the diffusion constant for that salt, $D_s=2D_1D_2/(D_1+D_2)$ \cite{newman2004}, where $D_1$ and $D_2$ are the individual ion diffusion coefficients (\Table{tab:ions}). This is simple to implement but approximate. In reality both the local electric fields generated by each salt, and the non-local field they generate, will interfere with the diffusion of both salts. We expect the results to be qualitatively the same with or without this effect. An example of a fully-coupled calculation is given in \Refcite{warren2020}
\item Once we have both salt profiles we combine them to compute the total salt concentration field, the conductivity $\sigma$ (and hence resitivity $\varrho$), the $g$ field, plus gradients of all three of these scalar fields.
\item Once we have $\grad g$ and $\grad \varrho$, we use \Eq{eq:source} in two dimensions to compute the (scalar) $S_z$. Then using $S_z$ we usea standard relaxation algorithm to solve \Eq{eq:2da} and hence obtain the potential $A_z$. Here we need boundary conditions along the four sides of our domain. We choose $A_z=0$ along these four sides.
\item Now that we have the potential $A_z$ we use numerical differentiation and \Eq{eq:2d} to obtain the current $\Ivec$. As the resistivity is known, we then easily calculate the electric field from $\Evec=\varrho\Ivec$.
\item Diffusiophoretic velocities are obtained from \Eq{eq:3DP}. The non-local velocities are given by the last term, and the local contribution is the sum of the first two terms.
\end{enumerate}
Note that the background salt is assumed to sodium chloride, while the salt beacons are sodium chloride and potassium acetate.

\section{Diffusiophoresis in multiple salt gradients}\label{app:3DP}
Here we summarise results from \Refscite{warren2020} and~\cite{williams2024} for the DP drift velocity in the presence of gradients of multiple salts. For simplicity we assume that all ionic species are univalent.  The extension to multivalent ions is straightforward if a bit tedious. If $c=\sum_ic_i$ is the total ion concentration, the essential result is
\begin{equation}
  \Uvec = 4\,\frac{\epsilon_r\epsilon_0}{\eta}\Bigl(\frac{\kT}{e}\Bigr)^2
  \ln\cosh\Bigl(\frac{e\zeta}{4\kT}\Bigr)\,\grad\ln c
  + \frac{\epsilon_r\epsilon_0\zeta}{\eta}\,\Evec\,.\label{eq:app:DP}
\end{equation}
The first term is chemiphoresis up the overall salt gradient independent of sign of the $\zeta$-potential, and the second term is electrophoresis in the electric field.  In the NP problem, the electric field can be decomposed as $\Evec=\varrho\grad g+\varrho\Ivec$.  Injecting this into \Eq{eq:app:DP} obtains \Eq{eq:3DP} in the main text with 
\begin{equation}
    \Gamma_0=\frac{\epsilon_r\epsilon_0}{\eta}
    \Bigl(\frac{\kT}{e}\Bigr)^2\,,\>
    \alpha=4\ln\cosh\Bigl(\frac{e\zeta}{4\kT}\Bigr)\,,\>
    \gamma=\frac{e\zeta}{\kT}\,,
\end{equation}
for the coefficients.

\section{Diffusio-osmosis driven by the walls}\label{app:do}
The two-dimensional systems that we envisage in section \ref{sec:2d} are contained in a gap between two plates.  As such, it seems inevitable that diffusio-osmosis (DO) will take place at the walls, conferring a flow field to the confined solution.  The observed drift of suspended particles is then a combination of DP according to the above models, and advection in the DO flow field.  In this Appendix we discuss this latter aspect.  Our starting point is \Eq{eq:3DP} for the DP drift of a suspended particle, which can be re-purposed to describe DO slip at a wall with a slip velocity given by 
\begin{equation}
  -\vslip/\Gamma_0=\alpha \grad\ln c+\gamma\varrho\grad g+\gamma\varrho\Ivec\,,\label{eq:vslip}
\end{equation}
where $\alpha$ and $\gamma$ are calculated using the $\zeta$-potentials of the walls.  The negative sign arises because in DO it is the fluid that moves and the wall that is stationary, whereas in DP it is the particle that moves and the fluid that is stationary \cite{dp-note}.  

This wall slip drives fluid motion in the gap between the top and bottom surfaces. If the two surfaces are identical \cite{do-note}, this flow is a combination of plug flow and two-dimensional Poiseuille flow. The plug flow is driven by the wall slip, while the pressure gradients driving Poiseuille flow come from the two-dimensional incompressibility constraint.  The theory here can be adapted from that of a Hele-Shaw cell \cite{faber1995, Gu2018}.  The vertically-resolved flow field is
\begin{equation}
  \vvec(x, y, z)=\vslip(x, y)-({1}/{2\eta})z(h-z)\grad p\,,
\end{equation}
where $z$ is the vertical direction, $h$ is the gap (assumed constant), and $\grad p$ is the two-dimensional gradient of the pressure field.  Vertically averaging this obtains $\vvec=\vslip-(h^2/12\eta)\grad p$.  The prefactor $h^2/12\eta$ cancels out in the end, so for notational simplicity we define $\vartheta=(h^2/(12\eta)p$ as a reduced pressure field.  The equations to be solved are then
\begin{equation}
  \vvec=\vslip-\grad\vartheta\,,\quad \div\vvec=0\,.\label{eq:vvec}
\end{equation}

We take the divergence of the first of \Eqs{eq:vvec} to obtain $\delsq\vartheta=\div\vslip$, which is a Poisson equation for the pressure field.  Formally the solution can be written using the two-dimensional Green's function already encountered in section \ref{subsec:ff2d} as
\begin{equation}
  \vartheta = ({2\pi})^{-1}{\textstyle\int}\ln|\rvec-\rvec'|\,
  [\div\vslip]_{\rvec'}\,\dA'.
  \label{eq:thetaGreen}
\end{equation}
where $[\div\vslip]_{\rvec'}$ indicates the value evaluated at $\rvec'$.

\subsection{Diffuiso-osmosis in the far field}
In the far field we can use a multipole expansion for $\vartheta$, as we did for the potential $\rho A_z$. As before we expand out the logarithm (here in \Eq{eq:thetaGreen}, and the coefficient of the leading order term is
\begin{equation}
    \vartheta_0\sim {\textstyle\int} \div\vslip\,\dA = {\textstyle\int} \vslip\cdot\nvec\,\ds \,,
\end{equation}
where we used Gauss' theorem to write the integral as the integral of dot product of the slip velocity and the outward surface (curve in two dimension) normal. This integral over the edge of a disc will be in the far field.

In far field, only the term proportional to $\Ivec$ in \Eq{eq:vslip} is relevant since the other two contributions depend on concentration gradients which we suppose vanish.  So in the far field, $\vslip \propto \Ivec$, as $\rho$ is a constant there. So for the above integral one can use Gauss' theorem again to obtain
\begin{equation}
  \vartheta_0\sim {\textstyle\int} \vslip\cdot\nvec\
  \propto {\textstyle\int} \Ivec\cdot\nvec\
  = {\textstyle\int} \div\Ivec\,\dA  =0 \,,
\end{equation}
because $\div\Ivec=0$ since $\Ivec$ is solenoidal. Essentially, as (in far field) slip velocity is solenoidal, the monopole term in multipole expansion for the pressure is zero. Note that this argument that the monopole term is zero for the pressure is not the same as the argument for the monopole term in $\rho A_z$ being zero, that relied on the specific structure of the source $\Svec$.

But just as with the potential $\rho A_z$, the leading non-zero term for $\vartheta$ is dipolar, with coefficients that are proportional to the $(x, y)$ moments of $\div\vslip$ in the region where the gradients are localised.  This term contributes as $-\grad\vartheta$ in \Eq{eq:vvec}, so that it generates a dipolar circulating planar Poiseuille flow with streamlines  similar to the ion current shown in \Fig{fig:farfield}.

So in the far field we have a dipolar slip velocity {\em and} a dipolar pressure. So the net flow field due DO, $\vvec$ in \Eq{eq:vvec}, is a superimposition of a plug-like dipolar flow field corresponding to $\vslip$ arising from the $\Ivec$ term in \Eq{eq:vslip}, and a Poiseuille-like dipolar flow field coming from the incompressibility constraint manifested as the $\grad\vartheta$ term.  For any given horizontal slice, the Poiseuille contribution will be maximal in the mid-plane, and the plug flow will dominate near the walls \cite{couette-note}, so that the net combination will be $z$-dependent, but still dipolar.

The net motion of suspended particles in far field will be DP in the electric field of the ion current, superimposed on convection by the DO flow field.  Since \emph{all} of these are dipolar in character in far field, the prediction is that the \emph{net} particle motion will again be dipolar, with particles following `effective' streamlines which will look like the ion current lines in \Fig{fig:farfield}.  However some smearing of this will arise from diffusion in the vertical direction, or in imaging when vertically averaging the particle trajectories. 


\begin{thebibliography}{46}
\bibitem [{nor()}]{norton-note}%
  \BibitemOpen
  \href@noop {} {}\bibinfo {note} {The Norton equivalent circuit
  (\partFig{fig:gradg}{b}) is preferred here since it decouples the gradient in
  $g$ from the gradient in $\varrho$.}\BibitemShut {Stop}%
\bibitem [{do-()}]{do-note}%
  \BibitemOpen
  \href@noop {} {}\bibinfo {note} {If the wall slip is different at the top and
  bottom, the vertically-resolved flow field would be a combination of plug
  flow, planar Couette flow, and planar Poiseuille flow. Vertically averaging
  would then yield the same governing equation with an effective $\vslip$ being
  the average of the top and bottom walls. Since these are individually driven
  by the same gradient terms in \Eq{eq:vslip}, one can proceed using the mean
  values of $\alpha$ and $\gamma$.}\BibitemShut {Stop}%
\bibitem [{cou()}]{couette-note}%
  \BibitemOpen
  \href@noop {} {}\bibinfo {note} {If the wall slip is different at the top and
  bottom, there will be an additional Couette contribution.}\BibitemShut
  {Stop}%
\end{thebibliography}
%

\end{document}